# Controlling subluminal to superluminal behavior of group velocity in an f-deformed Bose-Einstein condensate beyond the rotating wave approximation

Z. Haghshenasfard[1], M. Soltanolkotabi[2] and M. H. Naderi[3]

Quantum Optics Group, Department of Physics, University of Isfahan, 81746-73441, Isfahan, Iran

**Abstract**

In this paper, we investigate tunable control of the group velocity of a weak probe field propagating through an f-deformed Bose-Einstein condensate of $\Lambda$-type three-level atoms beyond the rotating wave approximation. For this purpose, we use an f-deformed generalization of an effective two-level quantum model of the three-level $\Lambda$-configuration without the rotating wave approximation in which the Gardiner's phonon operators for Bose-Einstein condensate are deformed by an operator-valued function, $f(\hat{n})$, of the particle-number operator $\hat{n}$. Corrections produced by the counter-rotating terms appear in the first order as an intensity-dependent detuning and in the second order as an intensity-dependent atom-field coupling. We consider the collisions between the atoms as a special kind of f-deformation where the collision rate $\kappa$ is regarded as the deformation parameter. We demonstrate the enhanced effect of subluminal and superluminal propagation based on electromagnetically induced transparency and electromagnetically induced absorption, respectively. In particular, we find that (i) the absorptive and dispersive properties of the deformed condensate can be controlled effectively in the absence of the rotating wave approximation by changing the deformation parameter $\kappa$, the total number of atoms $\hat{N}$ and the counter-rotating terms parameter $\lambda$, (ii) by increasing the values of $\lambda, \kappa$ and $\eta = 1/N$, the group velocity of the probe pulse changes, from subluminal to superluminal and (iii) beyond the rotating wave approximation, the subluminal and superluminal behaviors of the probe field are enhanced.

---

[1] E-mail: zhaghshenas@hotmail.com
[2] E-mail: soltan@sci.ui.ac.ir
[3] E-mail: mh.naderi@yahoo.com

# 1 Introduction

The manipulation of subluminal and superluminal light propagation in material systems has been the subject of growing interest over the years [1]. Normal dispersion leads to subluminal propagation and superluminal propagation corresponds to abnormal dispersion. Manipulation of light propagation in an optical medium can be realized by changing the dispersive properties of the medium. Experiments have recently demonstrated both subluminal [2, 3] and superluminal [4] propagation of light in an optical medium. Recently, tunable group velocity for light propagation from subluminal to superluminal in $\Lambda$-type and $V$-type three level atoms in a single system has been shown. For example, switching from subluminal to superluminal propagation can be achieved by changing the Rabi frequency of an additional microwave field [5], by increasing the intensity of the single driving field [6] and by adjusting the relative phase between probe and pump fields [7]. Experimental evidences of subluminal and superluminal propagation in four-level atoms have also been obtained [8]. The above-mentioned experiments of subluminal and superluminal propagation are performed in the electromagnetically induced transparency (EIT) and electromagnetically induced absorption (EIA), respectively. In 1999, Hau and coworkers [2] reported an experimental demonstration of EIT in a Bose-Einstein condensate (BEC) of sodium atoms, in which the optical pulses propagated at twenty million times slower than the speed of light in vacuum. As is known, to study the dynamics of the BEC gas, the Bogoliubov approximation [9] in quantum many body theory is usually applied, in which the creation and annihilation operators for condensate atoms are replaced by a c-number. However, this approximation destroys the conservation of the total particle number. To overcome this problem, Gardiner [10] suggested a modified Bogoliubov approximation by introducing phonon operators which conserve the total atomic particle number $N$ and obey the f- deformed commutation relation of the Heisenberg-Weyl algebra such that as $N \to \infty$, the usual commutation relation of the Heisenberg-Weyl algebra is regained. Then the Gardiner's phonon approach gives an elegant infinite atomic particle- number approximation theory for BEC taking into account the conservation of the total atomic number [11].

In the last few years, there have attracted tremendous interests in the development of studies for deformations of classical Lie algebras, so-called quantum groups, quantum algebras and quantum spaces, whose rich structures have produced important results in different areas of mathematics and physics. Quantum groups and quantum algebras have been applied to several research areas of physics and mathematics such as exactly solvable statistical models [12], non commutative geometry [13], nuclear quantum many body problems [14] and rational conformal field theories [15]. The study of quantum groups have led to the notion of q-deformed boson oscillators [16] based on deformation of the standard algebra of boson creation and annihilation operators. The q-deformed boson oscillator has been interpreted as a nonlinear oscillator with a very specific type of nonlinearity which classically corresponds to an intensity dependence of the oscillator frequency [17]. In addition, a general type of nonlinearity for which the intensity dependence of frequency of oscillations is described by a generic function $f(\hat{n})$, so-called f-deformation, has been introduced [18]. The relation between deformed radiation field and nonlinear quantum optical processes has been studied [19]. Recently, much attention has been paid to understanding and applications of the f-deformed bosons for the description of BEC [20]. The type of quantum nonlinearity introduced by f-deformation provides a compact description of physical effects in BEC. For example, the f-deformed BEC produces a correction to the Plank distribution formula [21, 22]. The statistical and thermodynamical properties of a gas of f-deformed bosons have been investigated [23].

BEC has commonly been referred to as Dicke model [24]. The quantum electrodynamical model involving the interaction of a single mode of the electromagnetic field with $N$ identical two-level atoms, known popularly as the Dicke model [24] in the literature, is of interest in many areas of physics. Many aspects of the Dicke model have been studied, such as critical behavior of the atom-field entanglement [25, 26], squeezing [27] and quantum phase transition [28]. A change from quasi-integrable to quantum chaotic behavior in the Dicke model has been established [29]. The finite-size corrections in the Dicke model have been studied [30]. These corrections have been shown to be crucial in the understanding of entanglement properties [26].

As we know, the Dicke model is analytically solvable due to the RWA, which is a popular technique in quantum optics, laser physics and other branches of resonance phenomena [31]. It leads to some mathematical simplifications in the calculation procedures of many problems. Under this approximation the rapidly oscillating counter-photon transitions are dropped from the interaction Hamiltonian. However, if the counter-rotating terms (CRTs) are considered, the models are not solvable by usual techniques since the eigenstates of the Hamiltonian can not be found in a closed form. In physical terms, this generalization allows us to introduce an additional parameter $\lambda$ into the Dicke model. The physical nature of the additional parameter $\lambda$ might be treated as some nonlinearity. The use of Dicke model beyond the RWA, in this case, corresponds to a phenomenological description of the nonlinear interaction of the atoms with the electromagnetic fields under the RWA. Although the RWA is a very good approximation, the CRTs contributions in the Dicke model have received increasing attention from many authors, since they can yield physical effects, albeit small ones. Influence of the CRTs on the population inversion in the Dicke model with cavity damping for an initial coherent field has been studied [32]. It has been shown [33] that the CRTs could lead to significant changes in the Rabi oscillations and squeezing in the Dicke model in the absence of cavity losses. Analogous investigations in cavity with losses [34] have confirmed the significance of the contribution of the CRTs, in which the effect of the CRTs leads to the appearance of new steady states. The role of the CRTs in the dynamical and steady state behavior of the dipole moment, mean photon number and squeezing parameters in the Dicke model with cavity losses have been examined [32]. All the studies reveal the significance of further investigation of quantum properties in the Dicke model with the CRTs included. Furthermore, it is reasonable to expect that the optical properties of an atomic medium and propagation properties of optical pulses in the medium are modified in the presence of CRTs.

In the present paper, we investigate tunable control of the group velocity of an optical pulse propagating through an f-deformed BEC beyond the RWA. The system under consideration is an f- deformed BEC of a gas with $\Lambda$- type three- level atoms in the EIT regime, in which the Gardiner's phonon operators for BEC are deformed by an operator-valued function $f(\hat{n})$. By considering the CRTs in the interaction Hamiltonian and the effect of collision between the atoms within the condensate as a special kind of the f-deformation, in which the collision rate $\kappa$, is regarded as a deformation parameter, we analyze the light propagation in the deformed BEC. Such a system offers extra degrees of flexibility $(\kappa, N, \lambda)$ for processing signal, and an effective control of the group velocity of light in deformed BEC can be demonstrated by changing the controlling parameters $(\kappa, N, \lambda)$ which leads to both subluminal and superluminal propagation of light. We show that the presence of both CRTs and f-deformation, exhibit nonlinear characteristics in the system under consideration, such that the nonlinearities may be increased by adjusting the controlling parameters $(\kappa, N, \lambda)$, and the nonlinearities enhance the subluminal and superluminal properties of light propagation.

The scheme of the paper is as follows. In section 2 we present our model and by using a method based on perturbation theory, we first show that in the absence of RWA, the model Hamiltonian can be transformed to an intensity-dependent Hamiltonian. Corrections produced by the CRTs appear in the first order as the intensity-dependent detuning and in the second order as the intensity-dependent atom-field coupling. In section 3 we use the deformed algebra to study the condensate with large but finite number of atoms. We show that a physical and natural realization of the f-deformed boson is provided by the Gardiner's phonon operators, for the description of the BEC. Here the deformation parameter is no longer phenomenological and is defined by the total number of atoms. We show that the effect of collisions between the atoms within condensate is an extra deformation on the intrinsically deformed Gardiner's phonon operators for BEC. In section 4 the quantum approach of the angular momentum is used to obtain the eigenvalues and eigenfunctions of the system up to first-order approximation. The interaction between the f-deformed BEC and the probe pulse and conditions for subluminality and superluminality of the probe pulse are studied in section 5. Finally we summarize our results in section 6.

## 2 The effective two- level model beyond the RWA

We consider the EIT effects in the BEC of $\Lambda$- type three- level atoms interacting with two laser fields (Fig. 1). The levels $|2\rangle$ and $|3\rangle$ are coupled by a probe field of amplitude $A_p$ and frequency $\omega_p$. The upper level $|3\rangle$ is coupled to level $|1\rangle$ by a classical coupling field of amplitude $A_c$ and frequency $\omega_c$. Under rotating wave approximation the total Hamiltonian for the atom and two fields is given by

$$\hat{H} = \hbar\omega_{12}|1\rangle\langle 1| + \hbar\omega|3\rangle\langle 3| + \hbar[(-g_1|3\rangle\langle 1|e^{-i\omega_c(t-\frac{z}{c})} - g_2|3\rangle\langle 2|e^{-i\omega_p(t-\frac{z}{c})}) + H.C], \qquad (2)$$

where coupling constants are defined by $g_1 = \dfrac{\mu_{31}A_c}{\hbar}$ and $g_2 = \dfrac{\mu_{32}A_p}{\hbar}$ with $\mu_{ij}$ denoting the transition dipole matrix element between states $|i\rangle$ and $|j\rangle$. By making the following substitutions,

$\rho_{ii} = \bar{\rho}_{ii}, \rho_{31} = \bar{\rho}_{31}e^{-i\omega_c(t-z/c)}, \rho_{32} = \bar{\rho}_{32}e^{-i\omega_p(t-z/c)} \rho_{12} = \bar{\rho}_{12}e^{-i(\omega_p-\omega_c)(t-z/c)}$.

the equations of motion for the atomic density matrix elements $\bar{\rho}_{ij}$ are given by

$$\dot{\bar{\rho}}_{11} = -ig_1\bar{\rho}_{13} + ig_1^*\bar{\rho}_{31} - 2\gamma_{12}\bar{\rho}_{11} + 2\gamma_{31}\bar{\rho}_{33},$$
$$\dot{\bar{\rho}}_{22} = -ig_2\bar{\rho}_{23} + ig_2^*\bar{\rho}_{32} + 2\gamma_{12}\bar{\rho}_{11} + 2\gamma_{32}\bar{\rho}_{33},$$
$$\dot{\bar{\rho}}_{33} = ig_1\bar{\rho}_{13} - ig_1^*\bar{\rho}_{31} + ig_2\bar{\rho}_{23} - ig_2^*\bar{\rho}_{32} - 2(\gamma_{32} + \gamma_{31})\bar{\rho}_{33},$$
$$\dot{\bar{\rho}}_{21} = -i(\delta - \Delta)\bar{\rho}_{21} - ig_1\bar{\rho}_{23} + ig_2^*\bar{\rho}_{31} - \gamma_{12}\bar{\rho}_{21},$$
$$\dot{\bar{\rho}}_{31} = -i\delta\bar{\rho}_{31} + ig_1(\bar{\rho}_{11} - \bar{\rho}_{33}) + ig_2\bar{\rho}_{21} - (\gamma_{12} + \gamma_{32} + \gamma_{31})\bar{\rho}_{31},$$
$$\dot{\bar{\rho}}_{32} = -i\Delta\bar{\rho}_{32} + ig_2(\bar{\rho}_{22} - \bar{\rho}_{33}) + ig_1\bar{\rho}_{12} - (\gamma_{32} + \gamma_{31})\bar{\rho}_{32},$$

(3)

where, $\Delta = \omega - \omega_p$, $\delta = \omega - \omega_{12} - \omega_c$ and the constants $\gamma_{ij}$ determine the rate of spontaneous decay from the level $|i\rangle$ to level $|j\rangle$ in the $\Lambda$- scheme. We consider the dynamics of the levels $|2\rangle$ and $|3\rangle$ of three- level $\Lambda$- type atoms interacting with the probe field in the EIT regime. We characterize the transition $|2\rangle \to |3\rangle$ in terms of the effective coupling constant associated with the dipole matrix element $\bar{\rho}_{32}$. Assuming all atoms being initially in the ground state $|2\rangle$ we obtain the following stationary solution

$$\bar{\rho}_{32} = \frac{1}{\Gamma} = [\Delta - 2i\gamma_{opt} + \frac{|g_1|^2}{i\gamma_{mag} - \Delta}]^{-1}, \tag{4}$$

where

$$\gamma_{opt} = \frac{\gamma_{31} + \gamma_{32}}{2}, \gamma_{mag} = \gamma_{12}. \tag{5}$$

We introduce the creation and annihilation operators $\hat{a}^+$ and $\hat{a}$, respectively, for the probe field and assume strong coupling field so that its intensity is given by a c- number parameter. Under the dipole approximation and in the absence of RWA the Hamiltonian (2) can be written as an effective two- level Hamiltonian in the following form

$$\hat{H} = \hat{H}_{free} + \hat{H}_{RWA} + \hat{H}_{CRT},$$
$$\hat{H}_{free} = \hbar\omega_p(\hat{a}^+\hat{a} + \hat{S}_3 + \frac{\hat{N}}{2}) + \Delta\hat{S}_3,$$
$$\hat{H}_{RWA} = \hbar k(\hat{a}\hat{S}_+ + \hat{a}^+\hat{S}_-),$$
$$\hat{H}_{CRT} = \hbar k(\hat{a}^+\hat{S}_+ + \hat{a}\hat{S}_-),$$

(6)

where $\hat{N}$ is the total number of atoms in BEC and the operators $\hat{S}_3$ and $\hat{S}_\pm$, describe total dipole momentum corresponding to the transitions $|3\rangle \to |2\rangle$ for the atoms in the BEC, $\hat{H}_{free}$ denotes the free energy of the probe field and atoms, $\hat{H}_{CRT}$ is counter rotating Hamiltonian and the atom-field coupling strength $k$ is defined by [35]

$$k = k_0 L_l, \tag{7}$$

where $k_0 = \mu_{32}\sqrt{\dfrac{\omega_p}{2\hbar\varepsilon_0 V}}$ is the single-photon Rabi-frequency in the Dicke model [24], and the parameter $L_l$ denotes linear coupling constant

$$L_l = \frac{\bar{\rho}_{32}(g_1,\Delta)}{\bar{\rho}_{32}(g_1=0,\Delta)}. \tag{8}$$

Here, by generalizing the perturbation theory used by Shirely [36] to the quantum field case, we obtain a simple effective Hamiltonian that takes into account the influence of CRTs. For this purpose, we approximately transform the Hamiltonian (6) to the one which commutes with the excitation number operator $\hat{N}' = \hat{a}^+\hat{a} + \hat{S}_3 = \hat{n} + \hat{S}_3$, by applying a sequence of small rotations. At the first step, we apply to the Hamiltonian (6) the transformation $\hat{U} = \exp[\lambda(\hat{a}^+\hat{S}_+ - \hat{a}\hat{S}_-)]$. Assuming $\lambda$ to be small, we neglect the terms of order $\lambda^j, j \geq 3$ and we obtain

$$\hat{U}\hat{H}\hat{U}^+ \approx \hat{H} - 2\hbar\lambda\bar{\omega}(\hat{a}^+\hat{S}_+ + \hat{a}\hat{S}_-) - \hbar\lambda^2\bar{\omega}[\hat{S}_3(2\hat{n}+1) - \hat{N}] + \hbar\lambda k(\hat{a}^{+2} + \hat{a}^2)\hat{S}_3 -$$
$$\hbar\lambda^2 k[(\hat{a}^{+3}\hat{S}_+ + \hat{a}^3\hat{S}_-) + (\hat{S}_+\hat{a}\hat{n} + \hat{n}\hat{a}^+\hat{S}_-)] + \hbar\lambda k[\hat{S}_3(2\hat{n}+1) - \hat{N}] - 2\hbar\lambda^2 k(\hat{S}_-\hat{a}\hat{n} + \hat{n}\hat{a}^+\hat{S}_+), \tag{9}$$

where $\bar{\omega} = (\omega_p + \Delta/2)$. It is obvious that if we choose $\lambda = k/2\bar{\omega}$ the second term in (9) cancels the counter rotating Hamiltonian $\hat{H}_{CRT} = k(\hat{a}^+\hat{S}_+ + \hat{a}\hat{S}_-)$. The transformed Hamiltonian $\hat{U}\hat{H}\hat{U}^+$ conserves the number of excitations $\hat{N}'$. It is noticeable that the parameter $\lambda = k/2\bar{\omega}$ is indeed small for a standard experimental setup in Ref [8].

Now, we perform the second small rotation $\hat{U}' = \exp[\dfrac{\lambda k}{2\omega_p}(\hat{a}^{+2} - \hat{a}^2)\hat{S}_3]$ to diagonalize the Hamiltonian up the second order in $\lambda$ in the basis of eigenstates $\hat{N}'$. Straightforward calculation shows that the Hamiltonian $\hat{H}$ after the transformations $\hat{U}$ and $\hat{U}'$ takes the following form

$$\hat{U}'\hat{U}\hat{H}\hat{U}^+\hat{U}'^+ \approx \hat{H}_{free} + \hat{H}_{RWA}^{(ID)} +$$
$$\hbar\lambda k[\hat{S}_3(\hat{n} + 1/2) - \hat{N}/2] - 2\hbar\lambda^2 k\frac{\bar{\omega}}{\omega_p}[(\hat{a}^3\hat{S}_+ + \hat{a}^{+3}\hat{S}_-) + \hat{\tilde{H}}, \tag{10}$$

where $\hat{H}_{RWA}^{(ID)}$ is the RWA interaction Hamiltonian with intensity-dependent coupling constant

$$\hat{H}_{RWA}^{(ID)} = \hbar k(\hat{n})\hat{a}^+\hat{S}_- + \hbar\hat{S}_+\hat{a}k(\hat{n}), \quad k(\hat{n}) = k(1 - \lambda^2\hat{n}), \tag{11}$$

and

$$\hat{\tilde{H}} = \hbar\lambda^2 k[\frac{(\omega_p + \Delta)}{\omega_p}(\hat{a}^{+3}\hat{S}_+ + \hat{a}^3\hat{S}_-) + \hbar\frac{\Delta}{\omega_p}(\hat{S}_+\hat{n}\hat{a}^+ + \hat{a}\hat{n}\hat{S}_-)]. \tag{12}$$

We now notice that $\hat{\tilde{H}}$ can be easily eliminated by applying corresponding transformations with rotation parameters proportional to $\lambda^3$ which, of course, does not affect the terms of the order of $\lambda^2$. In order to eliminate the first term in $\hat{\tilde{H}}$, i.e. $\hbar\lambda^2 k[\frac{(\omega_p+\Delta)}{\omega_p}(\hat{a}^{+3}\hat{S}_+ + \hat{a}^3\hat{S}_-)]$, describing the three-photon transitions we apply the transformation $\hat{U}'' \equiv \exp[\frac{2\overline{\omega}\lambda^2 k}{\omega_p[(\omega_p+\Delta)-3\omega_p]}(\hat{a}^3\hat{S}_+ - \hat{a}^{+3}\hat{S}_+)]$. If the atomic transition is far from the three-photon resonance, we get $\frac{2\overline{\omega}\lambda^2 k}{\omega_p[(\omega_p+\Delta)-3\omega_p]} \approx \lambda^3$, and the term can be eliminated. It is obvious that in the case of three-photon resonance $((\omega_p+\Delta)\approx 3\omega_p)$ the transformation $\hat{U}''$ can no longer be applied. Furthermore, it is easy to find the width of the three-photon resonance, $(\omega_p+\Delta)-3\omega_p \approx k\lambda^2$, which is much less than the width of the one-photon resonance, $(\omega_p+\Delta)-\omega_p \approx k$. Let us note that in the three-photon resonance case the term describing the one-photon transition, $k(\hat{a}\hat{S}_+ + \hat{a}^+\hat{S}_-)$, can be removed by applying appropriate small transformation.

Therefore the effective Hamiltonian for the case of one-photon resonance takes the following form

$$\hat{H}_{eff} = \hbar\{\omega_p(\hat{a}^+\hat{a}+\hat{S}_3+\frac{\hat{N}}{2}) + \Delta\hat{S}_3 + \lambda k[\hat{S}_3(\hat{n}+\frac{1}{2})-\frac{\hat{N}}{2}] + k(\hat{n})\hat{a}^+\hat{S}_- + \hat{S}_+\hat{a}k(\hat{n})\}. \qquad (13)$$

It is clear that $[\hat{H}_{eff}, \hat{N}'] = 0$. As is seen, the corrections produced by the CRTs appear in the first order as the intensity-dependent detuning, i.e., dynamical Stark shift, and in the second order as the intensity-dependent atom-field coupling. We can write the above Hamiltonian in the following form

$$\hat{H}_{eff} = \hbar\{\omega_p(\hat{a}^+\hat{a}+\hat{S}_3+\frac{\hat{N}}{2}) + \Delta\hat{S}_3 + \lambda k[\hat{S}_3(\hat{n}+\frac{1}{2})-\frac{\hat{N}}{2}] + \\ k(\hat{a}^+\hat{S}_- + \hat{S}_+\hat{a}) - \lambda^2 k(\hat{a}^+\hat{a}\hat{a}^+\hat{S}_- + \hat{S}_+\hat{a}\hat{a}^+\hat{a})\}, \qquad (14)$$

where, the first and the second terms give the free energy of the probe field and atoms respectively. The third term shows the intensity-dependent detuning. The fourth term represents linear contribution into the interaction between the field and two-level particles in dipole approximation. It has the typical form of the Dicke model. The last term describes nonlinear processes and it depends on the intensity of the probe pulse.

## 3 The f- deformed bosonic algebra for the Gardiner's phonon

We now apply the second quantization theory to the above model. Let $\hat{b}_{e(g)}, \hat{b}^+_{e(g)}$ denote the annihilation and creation operators of the atoms in the excited (ground) state, respectively, so that the total Hamiltonian reads as

$$\hat{H}_{eff} = \hbar[\omega_p \{\hat{a}^+\hat{a} + (\hat{b}_e^+\hat{b}_e - \hat{b}_g^+\hat{b}_g) + \frac{(\hat{b}_e^+\hat{b}_e + \hat{b}_g^+\hat{b}_g)}{2}\} + \Delta(\hat{b}_e^+\hat{b}_e - \hat{b}_g^+\hat{b}_g)$$
$$+ \lambda k[(\hat{b}_e^+\hat{b}_e - \hat{b}_g^+\hat{b}_g)(\hat{n} + \frac{1}{2}) - \frac{(\hat{b}_e^+\hat{b}_e + \hat{b}_g^+\hat{b}_g)}{2}] + k(\hat{a}^+\hat{b}_g^+\hat{b}_e + \hat{a}\hat{b}_e^+\hat{b}_g) \quad (15)$$
$$- \lambda^2 k(\hat{a}^+\hat{a}\hat{a}^+\hat{b}_g^+\hat{b}_e + \hat{a}\hat{a}^+\hat{a}\hat{b}_e^+\hat{b}_g)].$$

It should be noted that in the above Hamiltonian the total atom number $\hat{N} = \hat{b}_e^+\hat{b}_e + \hat{b}_g^+\hat{b}_g$ is conserved ($[\hat{N}, \hat{H}] = 0$). As well known, the central tool to study the dynamics of a BEC in the thermodynamic limit ($\hat{N} \to \infty$) is Bogoliubov approximation method [9], in which the condensate operators treated as $c$ numbers. One consequence of this is that the resulting approximate Hamiltonian does not conserve the total particle number ($[\hat{N}, \hat{H}_b] \neq 0$). In order to preserve the property of the initial model, we consider the Gardiner's phonon operators [10] defined by

$$\hat{b}_q = \frac{1}{\sqrt{N}}\hat{b}_g^+\hat{b}_e, \quad \hat{b}_q^+ = \frac{1}{\sqrt{N}}\hat{b}_g\hat{b}_e^+. \quad (16)$$

These operators obey a deformed algebra. In fact, a straightforward calculation leads to the following commutation relation

$$[\hat{b}_q, \hat{b}_q^+] = 1 - \frac{2}{N}\hat{b}_e^+\hat{b}_e = 1 - 2\eta\hat{b}_e^+\hat{b}_e, \quad (17)$$

By introducing the Gardiner's phonon operators we arrive at an intrinsically deformed model in which the deformation parameter is determined by the total number $N$. Having this in mind, the total Hamiltonian (15) can be written as [37]

$$\hat{H} = \hbar\omega_p\hat{a}^+\hat{a} + \hbar[\frac{3\omega_p}{2} + \Delta]\hat{b}_q^+\hat{b}_q +$$
$$\hbar\lambda k\hat{b}_q^+\hat{b}_q\hat{a}^+\hat{a} + \hbar K(\hat{a}^+\hat{b}_q + \hat{a}\hat{b}_q^+) - \hbar\lambda^2 K(\hat{a}^+\hat{a}\hat{a}^+\hat{b}_q + \hat{a}\hat{a}^+\hat{a}\hat{b}_q^+), \quad (18)$$

where $K = \sqrt{N}k$.

Now, we intend to study the effect of collisions between the atoms within condensate as a special kind of f-deformation. For this purpose, we remind the basics about the f-deformed quantum oscillators. The f-deformed quantum oscillators [22] by the algebra generated by the operators $\{1, \hat{A}, \hat{A}^+, \hat{N}\}$, where the operators $\hat{A}$ and $\hat{A}^+$ are defined as a nonlinear expansion of the usual harmonic oscillator operators $\hat{c}$ and $\hat{c}^+$,

$$\hat{A} = \hat{c}f(\hat{N}), \hat{A}^+ = f^*(\hat{N})\hat{c}^+, \hat{N} \equiv \hat{c}^+\hat{c}. \quad (19)$$

The operators $\hat{A}$ and $\hat{A}^+$ satisfy the following commutation relations

$$[\hat{A}, \hat{A}^+] = (\hat{N}+1)f^2(\hat{N}+1) - \hat{N}f^2(N) \quad [\hat{N}, \hat{A}] = -\hat{A} \quad [\hat{N}, \hat{A}^+] = -\hat{A}^+, \quad (20)$$

where, $f(\hat{N})$ is positive analytic function with $f(0) = 0$. The function $f(\hat{N})$ is specific to each f-deformed algebra. This function plays a central role in our treatment, since it determines the form of nonlinearities of the system under consideration. As a particular physical example, we consider the effect of collisions between the atoms within the condensate, where the operator valued function $f(\hat{N})$ is given as follows [37]

$$f_2(\hat{n}) = \sqrt{\kappa \hat{n} + (1-\kappa)}, \tag{21}$$

where the collision rate denoted by $\kappa$ and $\hat{n}$ is the particle number operator. The function $f_2(\hat{n})$, which is a characteristics for the deformation, has a dependence on the deformation parameter $\kappa$ such that when the deformation disappear (the case of no collision), then $f_2(\hat{n}, \kappa = 0) = 1$ and the usual algebra is recovered. In means the deformation increase with the collision rate $\kappa$. Subsequently, by considering the effect of collisions between the atoms within condensate, we can apply the extra deformation on the intrinsically deformed Gardiner's phonon operators for BEC by an operator- valued function $f_2(\hat{n}) = \sqrt{\kappa \hat{n} + (1-\kappa)}$ of the particle number operator $\hat{n}$ [37]. Here the nonlinearity is related to the collisions between the atoms within condensate. The deformed Gardiner's phonon operators are related to the nondeformed ones thorough the operator valued- function $f_2(\hat{n})$ as

$$\hat{B}_q = \hat{b}_q f_2(\hat{n}), B_q^+ = f_2^+(\hat{n}) \hat{b}_q^+,$$
$$\hat{n} = \hat{N}_q = \hat{b}_q^+ \hat{b}_q. \tag{22}$$

For small deformation, the f- deformed Gardiner's phonon operators $B_q, B_q^+$ can be expressed as

$$\hat{B}_q = \hat{b}_q[1 - \frac{\kappa}{2}(1 - \hat{b}_q^+ \hat{b}_q)], \quad \hat{B}_q^+ = [1 - \frac{\kappa}{2}(1 - \hat{b}_q^+ \hat{b}_q)]\hat{b}_q^+. \tag{23}$$

Here, the small value for $\kappa$ is considered, where $\kappa(1 - \hat{n})$ would be very smaller than one. In that case, by keeping only the lowest order of $\eta = \frac{1}{N}$ for very large total atom number $N$ and by keeping only the first- order term of the collision rate $\kappa$ for very low temperature we get

$$\hat{B}_q \approx (\hat{b} - \frac{1}{2N}\hat{b}^+\hat{b}\hat{b})[1 - \frac{\kappa}{2}\{1 - (\hat{b}^+\hat{b} - \frac{1}{N}\hat{b}^+\hat{b}^+\hat{b}\hat{b})\}],$$
$$\hat{B}_q^+ \approx [1 - \frac{\kappa}{2}\{1 - (\hat{b}^+\hat{b} - \frac{1}{N}\hat{b}^+\hat{b}^+\hat{b}\hat{b})\}](\hat{b}^+ - \frac{1}{2N}\hat{b}^+\hat{b}^+\hat{b}). \tag{24}$$

Therefore, by using (24) the deformed version of the Hamiltonian (18) can be expressed in terms of the nondeformed operators $\hat{b}$ and $\hat{b}^+$ as follows

$$\hat{H} = \hat{H}_0 + \hat{H}', \tag{25}$$

where

$$\hat{H}_0 = \hbar\omega_p(\hat{a}^+\hat{a} + \hat{b}^+\hat{b}) + \hbar K(\hat{a}\hat{b}^+ + \hat{a}^+\hat{b}), \tag{26}$$

and

$$\begin{aligned}\hat{H}' &= \hbar[\frac{\omega_p}{2}+\Delta](\hat{b}^+\hat{b}) + \hbar[\frac{3\omega_p}{2}+\Delta](-\frac{1}{N}+\kappa)\hat{b}^+\hat{b}^+\hat{b}\hat{b} \\ &+ \lambda k\hbar[\hat{b}^+\hat{b}\hat{a}^+\hat{a} + (-\frac{1}{N}+\kappa)\hat{b}^+\hat{b}^+\hat{b}\hat{b}\hat{a}^+\hat{a}] \\ &+ \hbar K(\frac{\kappa}{2}-\frac{1}{2N})(a\hat{b}^+\hat{b}^+\hat{b} + \hat{a}^+\hat{b}^+\hat{b}\hat{b}) + \\ &- \hbar\lambda^2 K[(\hat{a}\hat{a}^+\hat{a}\hat{b}^+ + \hat{a}^+\hat{a}\hat{a}^+\hat{b}) - \hbar\lambda^2 K(\frac{\kappa}{2}-\frac{1}{2N})(\hat{a}\hat{a}^+\hat{a}\hat{b}^+\hat{b}^+\hat{b} + \hat{a}^+\hat{a}\hat{a}^+\hat{b}^+\hat{b}\hat{b})].\end{aligned} \tag{27}$$

It is clear that the third and the last terms of $\hat{H}'$ are the corrections produced by the CRTs. Furthermore the second term in Eq. (27) describes the attractive exciton- exciton collisions due to bi- exciton effect. The third, the fourth and the last terms of $\hat{H}'$ describe the decrease of the exciton– photon coupling constants due to the phase- space filling effect [38]. It is evident that the Hamiltonian (27) reproduces the Hamiltonian (6) by setting $\lambda, \kappa$ and $\eta$ equal to zero.

## 4 Approximate analytical solutions

To solve the Schrödinger equation governed by the Hamiltonian (18) we shall make use of the quantum angular momentum theory. According to the Schwinger representation of the angular momentum, we can build the angular momentum operators as follows

$$\begin{aligned}\hat{J}_z &= \frac{1}{2}(\hat{a}^+\hat{a} - \hat{b}^+\hat{b}), \\ \hat{J}_+ &= \hat{a}^+\hat{b}, \\ \hat{J}_- &= \hat{a}\hat{b}^+.\end{aligned} \tag{28}$$

By using the cavity- field ladder operators $\hat{a}$ and $\hat{a}^+$ and the exciton operators $\hat{b}$ and $\hat{b}^+$ we define

$$\hat{J}_x = \frac{1}{2}(\hat{a}^+\hat{b} + \hat{a}\hat{b}^+), \hat{J}_y = \frac{1}{2i}(\hat{a}^+\hat{b} - \hat{a}\hat{b}^+). \tag{29}$$

Then we rewrite the Hamiltonian (26) as

$$\hat{H}_0 = \hbar\omega_p\hat{N} + 2\hbar K_1\hat{J}_x = \hbar\omega_p\hat{N} + 2\hbar K_1 e^{-i(\pi/2)\hat{J}_y}\hat{J}_z K_1 e^{+i(\pi/2)\hat{J}_y}, \tag{30}$$

The excitation number operator $\hat{N} = \hat{b}^+\hat{b} + \hat{a}^+\hat{a}$ is a constant under any SO(3) rotation and $\hat{J}^2 = \hat{J}_x^2 + \hat{J}_y^2 + \hat{J}_z^2 = \frac{\hat{N}}{2}(\frac{\hat{N}}{2}+1)$ is the total angular momentum operator. The common eigenstates of $\hat{J}^2$ and $\hat{J}_z$ are as follows

$$|jm\rangle = \frac{(\hat{b}^+)^{j-m}(\hat{a}^+)^{j+m}}{\sqrt{(j+m)!(j-m)!}}|0\rangle, \tag{31-a}$$

or

$$|n,n_e\rangle = \frac{(\hat{b}^+)^{n_e}(\hat{a}^+)^{n}}{\sqrt{n_e!n!}}|0\rangle, \tag{31-b}$$

where $n$ denotes the number of photons, $n_e$ is the number of excitons, and the eigenvalues of $J^2$ and $J_z$ are, respectively

$$j = \frac{N}{2}, m = -\frac{N}{2},...,\frac{N}{2}. \tag{32}$$

The eigenvectors $|\psi_{jm}^0\rangle$ and the eigenvalues $E_{jm}^{(0)}$ of $H_0$ can be easily constructed as

$$|\psi_{jm}^0\rangle = e^{-i(\pi/2)\hat{J}_y}|jm\rangle, \qquad E_{jm}^{(0)} = \hbar\omega_p N + 2\hbar K_1 m. \tag{33}$$

Up to the first-order approximation, the eigenvalues of the Hamiltonian $\hat{H}$ are obtained as

$$E_{jm} = E^{(0)}{}_{jm} + \langle jm|e^{i(\pi/2)\hat{J}_y}\hat{H}'e^{-i(\pi/2)\hat{J}_y}|jm\rangle, \tag{34}$$

and their corresponding eigenfunctions are given by

$$|\psi_{jk}\rangle = |\psi_{jk}^{(0)}\rangle + \sum \frac{\langle jn|\hat{H}'|jk\rangle}{E_{jk}^{(0)} - E_{jn}^{(0)}}|\psi_{jn}^{(0)}\rangle. \tag{35}$$

The matrix elements of the perturbation Hamiltonian $\hat{H}'$ are given by

$$\langle jm|e^{i(\pi/2)\hat{J}_y}\hat{H}'e^{-i(\pi/2)\hat{J}_y}|jm\rangle = \hbar(\frac{\omega_p}{2}+\Delta)[\frac{1}{2}(j+m)+\frac{1}{2}(j-m)]+$$

$$\hbar(\frac{3\omega_p}{2}+\Delta)(-\frac{1}{N}+\kappa)[\frac{1}{4}(j-m)(j-m-1)+\frac{1}{4}(j+m)(j+m-1)+(j^2-m^2)]+$$

$$\hbar\lambda k[\frac{1}{4}(j+m)^2+\frac{1}{4}(j-m)^2+\frac{1}{2}(j^2-m^2)-$$

$$\frac{1}{4}(j-m)(j+m-1)-\frac{1}{4}(j+m)(j-m-1)]+$$

$$\hbar\lambda k(-\frac{1}{N}+\kappa)\frac{1}{8}[(j+m)^2(j+m-1)+(j^2-m^2)(j+m-1)+$$

$$(j^2-m^2)(j-m-1)+(j-m)^2(j-m-1)-2(j^2-m^2)(j-m+1)-$$

$$2(j+m)(j+m-1)(j-m+1)-2(j-m)(j-m-1)(j+m+1)-$$

$$2(j^2-m^2)(j+m+1)+4(j-m)^2(j+m)+4(j+m)^2(j-m)]+$$

$$\hbar K(\frac{\kappa}{2}-\frac{1}{2N})[(\frac{1}{2}(j-m)(j-m-1)+\frac{1}{2}(j+m)(j+m-1)]-$$

$$\hbar\lambda^2 K[\frac{1}{2}(j+m)-\frac{1}{2}(j-m)+\frac{1}{2}(j+m)^2-\frac{1}{2}(j-m)^2+$$

$$(\frac{\kappa}{2}-\frac{1}{2N})\{\frac{1}{2}(j+m)(j+m-1)-\frac{1}{2}(j+m)(j+m-1)+$$

$$\frac{1}{4}(j+m)(j+m-1)(j+m-2)-\frac{1}{4}(j-m)(j-m-1)(j-m-2)-$$

$$\frac{1}{4}(j+m)(j-m)(j+m-1)+\frac{1}{4}(j+m)(j-m)(j-m-1)\}].$$
(36)

We shall apply this result in section 5 to drive the polarization of the medium.

## 5 Controlling group velocity in an f-deformed BEC

In order to get the susceptibility of the medium, we need to calculate the polarization of the atomic medium. To this end, we can make use of Hamiltonian approach [39]. According to this approach, the polarization of the medium can be expressed as the partial derivative of the averaged free energy density of the atomic medium with respect to the electric field amplitude

$$P = -\left\langle \frac{\partial \hat{H}'}{\partial E^*} \right\rangle,$$
(37)

where $\hat{H}'$ is the interaction part of the Hamiltonian, $E^*$ is the complex amplitude of the electromagnetic field and defined as $E = \varepsilon\sqrt{n}$, where $\varepsilon = \sqrt{\frac{\hbar\omega_p}{2\varepsilon_0 V}}$ ($V$ is the quantization volume) and $P$ is the polarization of the medium. Under adiabatic conditions [40], the averaged Hamiltonian that appears in Eq. (37) can be replaced by $E_{jm} - E^0_{jm}$. Then we get

$$P = -\left\langle \frac{\partial H'}{\partial E^*} \right\rangle = -\left\langle \frac{\partial(E_{jm} - E^0_{jm})}{\partial \varepsilon \sqrt{n}} \right\rangle =$$

$$= -\hbar\{(\frac{\omega_p}{2} + \Delta) + (\frac{3\omega_p}{2} + \Delta)(-\frac{1}{N} + \kappa)[-\frac{1}{2} + 2(j+m)] - \frac{\lambda k}{2} +$$

$$\lambda k(-\frac{1}{N} + \kappa)[\frac{1}{4}(j-m)^2 - \frac{1}{2}(j-m) + \frac{1}{2}] -$$

$$K(\frac{\kappa}{2} - \frac{1}{2N}) - \lambda^2 K - \lambda^2 K(\frac{\kappa}{2} - \frac{1}{2N})[\frac{1}{2}(j-m)^2]\}\sqrt{n} - \qquad (38)$$

$$\hbar\{(\frac{3\omega_p}{2} + \Delta)(-\frac{1}{N} + \kappa) + \lambda k + \lambda k(-\frac{1}{N} + \kappa)(\frac{(j-m)}{2} - \frac{3}{2}) +$$

$$2K(\frac{\kappa}{2} - \frac{1}{2N}) - 2\lambda^2 K + \lambda^2 K(\frac{\kappa}{2} - \frac{1}{2N})[1 + (j-m)]\}(\sqrt{n})^3 -$$

$$\hbar\{\frac{3}{2}\lambda k(\frac{\kappa}{2} - \frac{1}{2N}) - \frac{3}{2}\lambda^2 K(\frac{\kappa}{2} - \frac{1}{2N})\}(\sqrt{n})^5.$$

On the other hand, the total polarization of the BEC coupled to the probe electromagnetic field is given by

$$P = \varepsilon_0 \chi^{(1)}(\omega_p)(\varepsilon\sqrt{n}) + \varepsilon_0 \chi^{(3)}(\omega_p)(\varepsilon\sqrt{n})^3 + \varepsilon_0 \chi^{(5)}(\omega_p)(\varepsilon\sqrt{n})^5 + \ldots \qquad (39)$$

where $\chi^{(1)}$ is the linear susceptibility and $\chi^{(k)}$ represents the $k$ th-order nonlinear susceptibility. Using Eqs. (38) and (39) we obtain first, third and fifth order nonlinear susceptibilities of the BEC in the following form, (the higher order susceptibilities of the medium are zero).

$$\chi^{(1)}(\omega_p) = \frac{-\hbar}{\varepsilon^2 \varepsilon_0}\{(\frac{\omega_p}{2} + \Delta) + (\frac{3\omega_p}{2} + \Delta)(-\frac{1}{N} + \kappa)[-\frac{1}{2} + 2(j+m)] - \frac{\lambda k}{2} +$$

$$\lambda k(-\frac{1}{N} + \kappa)[\frac{1}{4}(j-m)^2 - \frac{1}{2}(j-m) + \frac{1}{2}] -$$

$$K(\frac{\kappa}{2} - \frac{1}{2N}) - K\lambda^2 - K\lambda^2(\frac{\kappa}{2} - \frac{1}{2N})[\frac{1}{2}(j-m)^2]\} +$$

$$\chi^{(3)}(\omega_p) = \frac{-\hbar}{\varepsilon^4 \varepsilon_0}\{(\frac{3\omega_p}{2} + \Delta)(-\frac{1}{N} + \kappa) + \lambda k + \lambda k(-\frac{1}{N} + \kappa)(\frac{(j-m)}{2} - \frac{3}{2}) + \qquad (40)$$

$$2K(\frac{\kappa}{2} - \frac{1}{2N}) - 2K\lambda^2 + K\lambda^2(\frac{\kappa}{2} - \frac{1}{2N})[1 + (j-m)],$$

$$\chi^{(5)}(\omega_p) = \frac{-\hbar}{\varepsilon^6 \varepsilon_0}\{\frac{3}{2}k\lambda(\frac{\kappa}{2} - \frac{1}{2N}) - \frac{3}{2}K\lambda^2(\frac{\kappa}{2} - \frac{1}{2N})\}.$$

The total susceptibility of the BEC at the probe field frequency including linear and nonlinear terms, reads

$$\chi = \chi^{(1)}(\omega_p) + \chi^{(3)}(\omega_p)|E(\omega_p)|^2 + \chi^{(5)}(\omega_p)|E(\omega_p)|^4 = \chi^{(1)}(\omega_p) + \chi^{(nl)}(\omega_p), \qquad (41)$$

where the nonlinear part of the total susceptibility of the medium is

$$\chi^{(nl)}(\omega_p) = \chi^{(3)}(\omega_p)|E(\omega_p)|^2 + \chi^{(5)}(\omega_p)|E(\omega_p)|^4. \qquad (42)$$

The refractive index $n(\omega_p)$ is related to the susceptibility of the medium $\chi(\omega_p)$ through the relation $n(\omega_p) = \sqrt{1 + \chi(\omega_p)}$, while the group velocity is $\frac{c}{n_g}$, where the group refraction index is defined as $n_g = n(\omega_p) + \omega_p \frac{dn(\omega_p)}{d\omega_p}$.

We are now in a position to present numerical results[Figs. 2-8] for dependence of the refraction group index and the real and imaginary parts of the linear and nonlinear susceptibilities of the f- deformed BEC under consideration on the CRTs (or $\lambda$ parameter) and the deformation parameters $\kappa$, $N$ (or $\eta = \frac{1}{N}$). In this paper, we consider probe pulse propagation in an f-deformed BEC of a gas of $\Lambda$-type configuration. This system can be realized in the BEC of sodium atoms (Fig. 1), where the probe transition is denoted by $\frac{\omega}{2\pi} = 5.1 \times 10^{14} Hz (\lambda = 589 nm)$ [2] and the energy splitting between the levels $|1\rangle$ and $|2\rangle$ is $\frac{\omega_{12}}{2\pi} = 1772 MHz$. The intensity of the coupling field is $I_c = \frac{55 mW}{cm^2}$ corresponding to the Rabi frequency $g_1/2\pi = 21.4 MHz$ [2], much stronger than the intensity of the probe field $I_p = \frac{80 \mu W}{cm^2}$ corresponding to 25 photons in average. The other physical parameters for the BEC of sodium are set to be, the density of the condensate $\frac{N}{V} = 3.3 \times 10^{12} cm^{-3}$ [2], the dipole matrix element $|\mu_{32}| = 22 \times 10^{-30} C.m$ [40], the decay rates $\gamma_{31}$ and $\gamma_{32}$ of the level $|3\rangle$ $\gamma_{31}/2\pi = \gamma_{32}/2\pi = 5 MHz$ [21], the decay rates from the transition between the hyperfine levels $|1\rangle$ and $|2\rangle$ $\gamma_{12}/2\pi = 38 KHz$ [41].

To observe the effects of CRTs on propagation of the probe pulse, for a given collision rate $\kappa$ and the total number of atoms $N$, we need to plot the group refraction index as a function of detuning $\Delta$, for different values of $\lambda$. This is shown in Fig 2a. The point $\Delta = 0$ of the exact resonance corresponds to the EIT regime, characterized by low losses. The zero points of group refraction index in Fig .3a show an uncertainty of the group velocity $v_g$, of the probe pulse and describe the possibility of observing superluminal velocities. When the refraction index is negative (corresponding to the superluminal propagation) the peak of the pulse exits the BEC before it passes the entrance face. We see that the pulse propagation changes from subluminal to superluminal and the significant subluminality and superluminality take place for large parameter $\lambda$. However, the increase in parameter $\lambda$ leads to large nonlinearity as shown in Fig 2b. In Fig 2b, we show the real part of the nonlinear total susceptibility $\chi^{(nl)}$, defined by Eq. (42), versus the probe pulse detuning $\Delta$, for three different values of the parameter $\lambda$, which correspond to the nonlinear properties of the medium. It is clear that the nonlinearity increases with the parameter $\lambda$ which in turn leads to the slower subluminal and faster superluminal group velocity. Therefore, for a given collision rate $\kappa$ and the total number of atoms $N$, the parameter $\lambda$ can be adjusted to realize the transition from subluminal to superluminal propagation. In Fig. 3a, we show the behavior of the group refraction index as a function of the detuning parameter $\Delta$ of the probe pulse for three different values of the deformation parameter $\kappa$ when $\lambda = 0.1$ and $N = 10^{20}$, $(\eta \approx 0)$. As is seen, with the increasing value of the collision rate $\kappa$ the subluminal and superluminal behaviors of the group velocity are enhanced. The physical origin of this result is that with increasing the deformation parameter $\kappa$ the nonlinearity of the model under consideration becomes larger, as shown in Fig. 3b. In Fig. 3b, we plot the real part of the nonlinear total susceptibility $\chi^{(nl)}$ as a function of $\Delta$ for three different values of $\kappa$ and $\lambda = 0.1$. The real part of $\chi^{(nl)}$ shows the nonlinearity of this process. It is evident that the nonlinearity increases with deformation parameter $\kappa$. To observe the effect of number of atoms $N$, for a given collision rate, in the presence of CRTs, ($\lambda = 0.1$) we plot the group refraction index as a function of detuning $\Delta$ for different values of $N$ (Fig. 4a). We see that the pulse propagation can change from subluminal to superluminal and significant superluminality and subluminality take place for large values of the deformation parameter $\eta$. In other words, superluminality and subluminality decrease with the total number of condensate atoms $N$. The reason is due to the fact that with increasing the deformation parameter $\eta$ there exist an enhancement of the nonlinearity as shown in Fig. 3b. In Fig. 4b, we plot the real part of the total nonlinear susceptibility $\chi^{(nl)}$ as a function of detuning $\Delta$ for different values of the deformation parameter $\eta$ and for the case of no collision ($\kappa = 0$) and $\lambda = 0.1$. As is seen, the nonlinearity increases with the deformation parameter $\eta$. The imaginary part of the total nonlinear susceptibility $\chi^{(nl)}$ is plotted versus the detuning $\Delta$ in Figs. 5a, 5b and 5c. We observe that the absorption coefficients increase with the parameters $\eta, \kappa$ and $\lambda$. In Figs. 6a, 6b and 6c, we plot the real part of the linear susceptibility $\chi^{(1)}$ versus detuning $\Delta$ for various values of the parameters $\eta, \kappa$ and $\lambda$. It is evident that the dispersive properties of the BEC increase with the parameter $\eta, \kappa$ and $\lambda$. The imaginary part of the linear susceptibility as a function of the detuning $\Delta$ is shown in Figs. 7a, 7b and 7c, for various values of the parameter $\eta, \kappa$ and $\lambda$. It is clear that the medium has absorption for positive absorption coefficient and gain for negative absorption coefficient. We see that the gain and absorption properties of the BEC increase with the parameters $\eta, \kappa$ and $\lambda$. In Figs. 8a-8f we illustrate the effects of the parameter $\lambda$ and deformation parameters $\eta, \kappa$ on the magnitude of the real and imaginary parts of the total susceptibility, which correspond to the dispersive and absorptive properties of the condensate, respectively. It is seen that at the region around the zero detuning $\Delta$ both $\chi', \chi''$ are equal to zero. This means that the absorption is almost zero where the index of refraction is unity. Thus the medium becomes transparent under the action of the probe field and EIT is occurred. It is evident that with the increasing values of the parameters $\eta, \kappa$ and $\lambda$ there exists an enhancement of the dispersion and absorption properties of the deformed BEC under consideration.

## 6 Summary and conclusions

In summary, we have studied the dispersive, absorptive and propagation properties of a weak optical probe field in an f- deformed BEC of the gas of $\Lambda$- type three- level atoms in the EIT regime beyond the RWA. By applying an effective two- level quantum model beyond the RWA, within the framework of the f- deformed boson model, we have shown that in the existence of CRTs, the model can be transformed into an intensity-dependent Hamiltonian. Corrections produced by the CRTs appear in the first order as the intensity-dependent detuning and in the second order as the intensity-dependent atom-field coupling. We have calculated the linear and nonlinear susceptibilities of the atomic condensates. In addition, we have considered the effects of collisions between the atoms within the condensate as a special kind of f- deformation for which the collision rate $\kappa$ is regarded as the corresponding deformation parameter. We have demonstrated that tunable control of the group velocity of a weak probe field from subluminal to superluminal and we have studied the effects of CRTs on the propagation of the probe pulse. We have found that the deformed parameters $\eta$, $\kappa$ and $\lambda$ play an important role in determining the subluminal and superluminal propagation through the condensate. We have also shown that by applying the deformation on atomic operators of the BEC medium, it is possible to obtain large nonlinearity that leads to an enhanced subluminal and superluminal propagation.

**References**

[1] S. E. Harris, Phys. Today **50**, 36 (1997); M. D. Lukin and A. Imamoglu, Nature **413**, 273 (1990); P. W. Milonni, J. Phys. B: At. Mol. Opt. Phys **35**, R31 (2002).

[2] L.V. Hau, S. E. Harris, Z. Dutton and C.H. Behroozi, Nature (London) **397**, 594 (1999 ).

[3] D. F. Phillips, A. Fleischhauer, A. Mair, R. L. Walswoth and M. D. Lukin, Phys. Rev. Lett. **86**, 783(2001); A. V. Turukhin, V. S. Sudarshanam, M. S. Shahriar, J. A. Musser, B. S. Hamm and P. R. Hemmer, Phys. Rev. Lett. **88**, 023602 (2002); O. Schmidt, R. Wynands, Z. Hussein and D. Meschede, Phys. Rev. A **53**, R27 (1996); D. Budker, D. F. Kimball, S. Rochester and V. V. Yashchuk, Phys. Rev. Lett. **83**, 1767 (1999).

[4] L.J. Wang, A. Kuzmich and A. Dogariu, Nature (London) **407**, 277 (2000); A. M. Akulshin, S. Barreiro and A. Lezama, Phys. Rev. Lett. **83**, 4277 (1999); A. M. Akulshin, A. Cimmino, A. I. Sidorov, P. Hannaford and G. I. Opat, Phys. Rev. A **67**, 011801 (2003); K. Kim, H. S. Moon, C. Lee, S. K. Kim, X. Z. Cheng, J. B. Kim, J. Phys. B: At. Mol. Opt. Phys **36**, 2671 (2003).

[5] G. S. Agarwal, T. N. Dey and S. Menon, Phys. Rev. A **64**, 053809 (2001).

[6] A. D. Wilson- Gordon and H. Friedmann, J. Mod. Opt. **49** 125 (2002).

[7] H. Sun, H. Guo, Y. Bai, D. Han, S. Fan and X. Chen, Phys. Lett. A **335**, 68 (2005).

[8] H. Kang, L. Wen and Y. Zhu, Phys. Rev. A **68**, 063806 (2003); F. Xiao, H. Guo, L. Li, C. Liu and X. Chen, Phys. Lett. A **327**, 15 (2004).

[9] N. N. Bogoliubov, J. Phys. (USSR) **11**, 23, (1947); reprinted in D. Pines, *the many body problem* (Benjamin, New York, 1962).

[10] C. W. Gardiner, Phys. Rev. A **56**, 1414(1997).

[11] X. X. Liu, C. P. Sun, S. X. Yu and D. L. Zhou, Phys. Rev. A **63**, 023802(2001).

[12] V. Pasquier, H. Saleur, Nucl. Phys. B **330**, 523 (1990).

[13] J. Wess, B. Zumino, Nucl. Phys. B **18**, 302 (1990).

[14] K. D. Sviratcheva, C. Bahri, A. I. Georgieva, J. P. Draayer, Phys. Rev. Lett. **93**, 152501 (2004).

[15] L. Alvarez-Gaume, C. Gomez and G. Sierra, Nucl. Phys. B **330**, 347 (1990).

[16] L. Biedenharn, J. Phys. A **22**, L873 (1989).

[17] V. I. Man'ko, G. Marmo, S. Solimeno and F. Zaccaria, Int. J. Mod. Phys. A **8**, 3577 (1993).

[18] V. I. Man'ko, G. Marmo, S. Solimeno and F. Zaccaria, Phys.Lett. A **176**, 173(1993); R. L. Matos Filho and W. Vogel, Phys. Rev. A **54**, 4560 (1996).

[19] M. H. Naderi, M. Soltanolkotabi and R. Roknizadeh, J. Phys. Soc. Japan , **73**, 2413 (2004).

[20] K. B. Davis et al., Phys. Rev. Lett. **75**, 3969(1995); G. Su, J. Chen and L. Chen, J. Phys. A: Math. Gen. **36** (2003); C. C. Bradley, C. A. Sackett, J. J. Tollett and R. G. Hulet, Phys. Rev. Lett. **75**, 1687(1995); A. Alighn and B. Deviren, J. Phys. A: Math. Gen. **38** (2005).

[21] P. Aniello, V. I. Man'ko, G. Marmo, S. Solimeno and F. Zaccaria, J. Opt. B: Quant. Semiclass. Opt. 2, 718 (2000).

[22] V. I. Man'ko, G. Marmo, E. C. G. Sudarshan and F. Zaccaria, Phys. Scripta **55**, 528 (1997).

[23] J.A. Tuszynski, J.L. Rubin, J. Meyer, M. Kibler, Phys. Lett. A **175**, 173 (1993).

[24] R. Dicke, Phys. Rev. 93, **99** (1954); M. Tavis and F. W. Cumming, Phys. Rev. **170**, 379(1968).

[25] N. Lambert, C. Emary and T. Brandes, Phys. Rev. Lett. **92**, 073602 (2004).

[26] N. Lambert, C. Emary and T. Brandes, Phys. Rev. A **71**, 053804 (2005).

[27] G. Ramon, C. Brif and A. Mann, Phys. Rev. A **58**, 2506 (1998); B. W. Shore and P. L. Knight, J. Mod. Opt. **40**, 1195 (1993).

[28] C. Emary and T. Brandes, Phys. Rev. Lett. **90**, 044101 (2003).

[29] G. A. Finney, J. Gea-Banacloche, Phys. Rev. E **54**, 1449 (1996).

[30] J. Vidal and S. Dusuel, Europhys, Lett. **74**, 817 (2006)

[31] L. Allen and J. H. Eberly, *Optical Resonance and Two-level atoms* (Willey, New York, 1975).

[32] J. Seke, Physica A, **240**, 635 (1997).

[33] J. Seke, Physica A, **213**, 587 (1995); J. Seke, Quant. Opt, **3**, 127 (1991).

[34] J. Seke, Physica A, **193**, 587 (1993).

[35] S. M. Arakelyan and A. V. Prokhorov, Phys. Rev. A **72**, 013804 (2005).

[36] J. H. Shirley, Phys. Rev. **138**, B979 (1965).

[37] Z. Haghshenas, M. H. Naderi and M. Soltanolkotabi (submitted).

[38] Jun- ichi Inoue, Tobias Brandes and Akira Shimizu, J. Phys. Soc. Jpn. **67**, 3384 (1998).

[39] B. Kryzhanovsky and B. Glushko, Phys. Rev. A **45**, 4979 (1992).

[40] M. Johnsson and M. Fleischhauer, Phys. Rev. A **67**, 061802 (2003).

[41] S. Al- Awfi and M. Babiker, Phys. Rev. A. **58**, 4763 (2000).

**Figure captions**

Fig.1: The energy level $\Lambda$ scheme of three-level atoms (for example, $^{23}Na$ atoms). Two hyperfine sub-levels of sodium state $3^2S_{1/2}$ with F=1 and F=2 are associated with levels $|2\rangle$ and $|1\rangle$ of the $\Lambda$- scheme, respectively. An excited state $|3\rangle$ corresponds to the hyperfine sub-level of the term $3^2P_{3/2}$ with F=2

Fig.2a: The group velocity as a function of the detuning parameter $\Delta$, for three different values of the parameter $\lambda$, $\lambda = 0(...)$, $\lambda = 0.1(-)$, $\lambda = 0.3(-.-.)$, the total number of atoms $N = 10^{20}$, $(\eta \approx 0)$ and the total number of excited atoms $n_e = 1$

Fig.2b: The real part of the total nonlinear susceptibility $\text{Re}(\chi^{(nl)})$, as a function of the detuning parameter $\Delta$, for three different values of the parameter $\lambda$, $\lambda = 0(...)$, $\lambda = 0.1(-)$, $\lambda = 0.3(-.-.)$, the total number of atoms $N = 10^{20}$, $(\eta \approx 0)$ and the total number of excited atoms $n_e = 1$

Fig.3a: The group velocity as a function of the detuning parameter $\Delta$, for three different values of the deformation parameter $\kappa$, $\kappa = 0(...)$, $\kappa = 0.005\text{Hz}(-)$, $\kappa = 0.008\text{Hz}(-.-.)$, the total number of atoms $N = 10^{20}$, $(\eta \approx 0)$, the total number of excited atoms $n_e = 1$ and $\lambda = 0.1$

Fig.3b: The real part of the total nonlinear susceptibility $\text{Re}(\chi^{(nl)})$, as a function of the detuning parameter $\Delta$, for three different values of the deformation parameter $\kappa$, $\kappa = 0(...)$, $\kappa = 0.005\text{Hz}(-)$, $\kappa = 0.008\text{Hz}(-.-.)$, the total number of atoms $N = 10^{20}$, $(\eta \approx 0)$, the total number of excited atoms $n_e = 1$ and $\lambda = 0.1$

Fig.4a: The group velocity as a function of the detuning parameter $\Delta$, for three different values of the total number of atoms $N$, $N = 30(...)$, $N = 20(-)$, $N = 10(-.-.)$ in the case of no collision ($\kappa = 0$), the total number of excited atoms $n_e = 1$ and $\lambda = 0.1$

Fig.4b: The real part of the total nonlinear susceptibility $\text{Re}(\chi^{(nl)})$, as a function of the detuning parameter $\Delta$, for three different values of the total number of atoms $N$, $N = 30(...)$, $N = 20(-)$, $N = 10(-.-.)$ in the case of no collision ($\kappa = 0$), the total number of excited atoms $n_e = 1$ and $\lambda = 0.1$

Fig.5a: The imaginary part of the total nonlinear susceptibility $\text{Im}(\chi^{(nl)})$, as a function of the detuning parameter $\Delta$, for three different values of the parameter $\lambda$,

$\lambda = 0(...)$, $\lambda = 0.1(-)$, $\lambda = 0.3(-.-.)$, the total number of atoms $N = 10^{20}$, $(\eta \approx 0)$ and the total number of excited atoms $n_e = 1$

Fig.5b: The imaginary part of the total nonlinear susceptibility $\text{Im}(\chi^{(nl)})$, as a function of the detuning parameter $\Delta$, for three different values of the deformation parameter $\kappa$, $\kappa = 0(...)$, $\kappa = 0.005\text{Hz}(-)$, $\kappa = 0.008\text{Hz}(-.-.)$, the total number of atoms $N = 10^{20}$, $(\eta \approx 0)$, the total number of excited atoms $n_e = 1$ and $\lambda = 0.1$

Fig.5c: The imaginary part of the total nonlinear susceptibility $\text{Im}(\chi^{(nl)})$, as a function of the detuning parameter $\Delta$, for three different values of the total number of atoms $N$, $N = 30(...)$, $N = 20(-)$, $N = 10(-.-.)$ in the case of no collision ($\kappa = 0$), the total number of excited atoms $n_e = 1$ and $\lambda = 0.1$

Fig.6a: The real part of the linear susceptibility $\text{Re}(\chi^{(1)})$ as a function of the detuning parameter $\Delta$, for three different values of the parameter $\lambda$, $\lambda = 0(...)$, $\lambda = 0.1(-)$, $\lambda = 0.3(-.-.)$, the total number of atoms $N = 10^{20}$, $(\eta \approx 0)$ and the total number of excited atoms $n_e = 1$

Fig.6b: The real part of the linear susceptibility $\text{Re}(\chi^{(1)})$ as a function of the detuning parameter $\Delta$, for three different values of the deformation parameter $\kappa$, $\kappa = 0(...)$, $\kappa = 0.005\text{Hz}(-)$, $\kappa = 0.008\text{Hz}(-.-.)$, the total number of atoms $N = 10^{20}$, $(\eta \approx 0)$ the total number of excited atoms $n_e = 1$ and $\lambda = 0.1$

Fig.6c: The real part of the linear susceptibility $\text{Re}(\chi^{(1)})$ as a function of the detuning parameter $\Delta$, for three different values of the total number of atoms $N$, $N = 300(...)$, $N = 200(-)$, $N = 100(-.-.)$ in the case of no collision ($\kappa = 0$), the total number of excited atoms $n_e = 1$ and $\lambda = 0.1$

Fig.7a: The imaginary part of the linear susceptibility $\text{Im}(\chi^{(1)})$ as a function of the detuning parameter $\Delta$, for different values of the parameter $\lambda$, $\lambda = 0(...)$, $\lambda = 0.1(-)$, $\lambda = 0.3(-.-.)$, the total number of atoms $N = 10^{20}$, $(\eta \approx 0)$ and the total number of excited atoms $n_e = 1$

Fig.7b: The imaginary part of the linear susceptibility $\text{Im}(\chi^{(1)})$ as a function of the detuning parameter $\Delta$, for three different values of the deformation parameter $\kappa$, $\kappa = 0(...)$, $\kappa = 0.005\text{Hz}(-)$, $\kappa = 0.008\text{Hz}(-.-.)$, the total number of atoms $N = 10^{20}$, $(\eta \approx 0)$ the total number of excited atoms $n_e = 1$ and $\lambda = 0.1$
.

Fig.7c: The imaginary part of the linear susceptibility $\text{Im}(\chi^{(1)})$ as a function of the detuning parameter $\Delta$, for three different values of the total number of atoms $N$, $N=30(...), N=20(-), N=10(-.-.)$ in the case of no collision ($\kappa=0$) the total number of excited atoms $n_e=1$ and $\lambda=0.1$

Fig.8a: The real part of the susceptibility $\text{Re}(\chi)$, as a function of the detuning parameter $\Delta$, for different values of the parameter $\lambda$, $\lambda=0(...), \lambda=0.1(-), \lambda=0.3(-.-.)$, the total number of atoms $N=10^{20}, (\eta\approx 0)$ and the total number of excited atoms $n_e=1$

Fig.8b: The real part of the susceptibility $\text{Re}(\chi)$, as a function of the detuning parameter $\Delta$, for three different values of the deformation parameter $\kappa$, $\kappa=0(...), \kappa=0.005\text{Hz}(-), \kappa=0.008\text{Hz}(-.-.)$, the total number of atoms $N=10^{20}, (\eta\approx 0)$ the total number of excited atoms $n_e=1$ and $\lambda=0.1$

Fig.8c: The real part of the susceptibility $\text{Re}(\chi)$, as a function of the detuning parameter $\Delta$, for three different values of the total number of atoms $N$, $N=30(...), N=20(-), N=10(-.-.)$ in the case of no collision ($\kappa=0$), the total number of excited atoms $n_e=1$ and $\lambda=0.1$

Fig.8d: The imaginary part of the susceptibility $\text{Im}(\chi)$, as a function of the detuning parameter $\Delta$, for different values of the parameter $\lambda$, $\lambda=0(...), \lambda=0.1(-), \lambda=0.3(-.-.)$, the total number of atoms $N=10^{20}, (\eta\approx 0)$ and the total number of excited atoms $n_e=1$

Fig.8e: The imaginary part of the susceptibility $\text{Im}(\chi)$, as a function of the detuning parameter $\Delta$, for three different values of the deformation parameter $\kappa$, $\kappa=0(...), \kappa=0.005\text{Hz}(-), \kappa=0.008\text{Hz}(-.-.)$, the total number of atoms $N=10^{20}, (\eta\approx 0)$ the total number of excited atoms $n_e=1$ and $\lambda=0.1$

Fig.8f: The imaginary part of the susceptibility $\text{Im}(\chi)$, as a function of the detuning parameter $\Delta$, for three different values of the total number of atoms $N$, $N=30(...), N=20(-), N=10(-.-.)$ in the case of no collision ($\kappa=0$) the total number of excited atoms $n_e=1$ and $\lambda=0.1$

Fig. 1

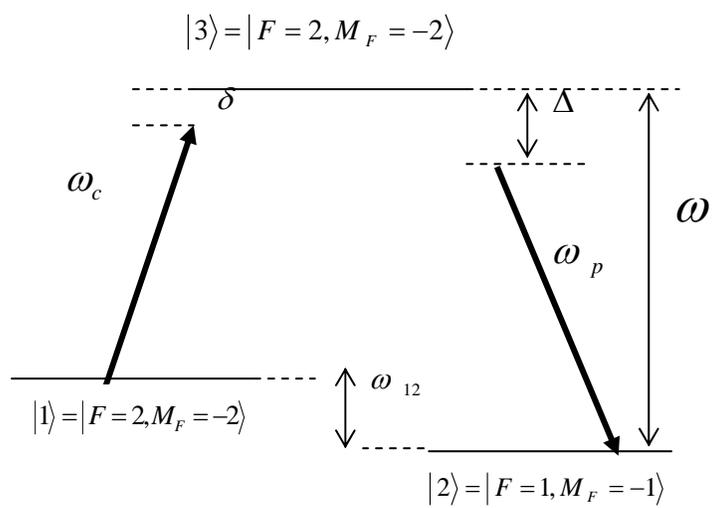

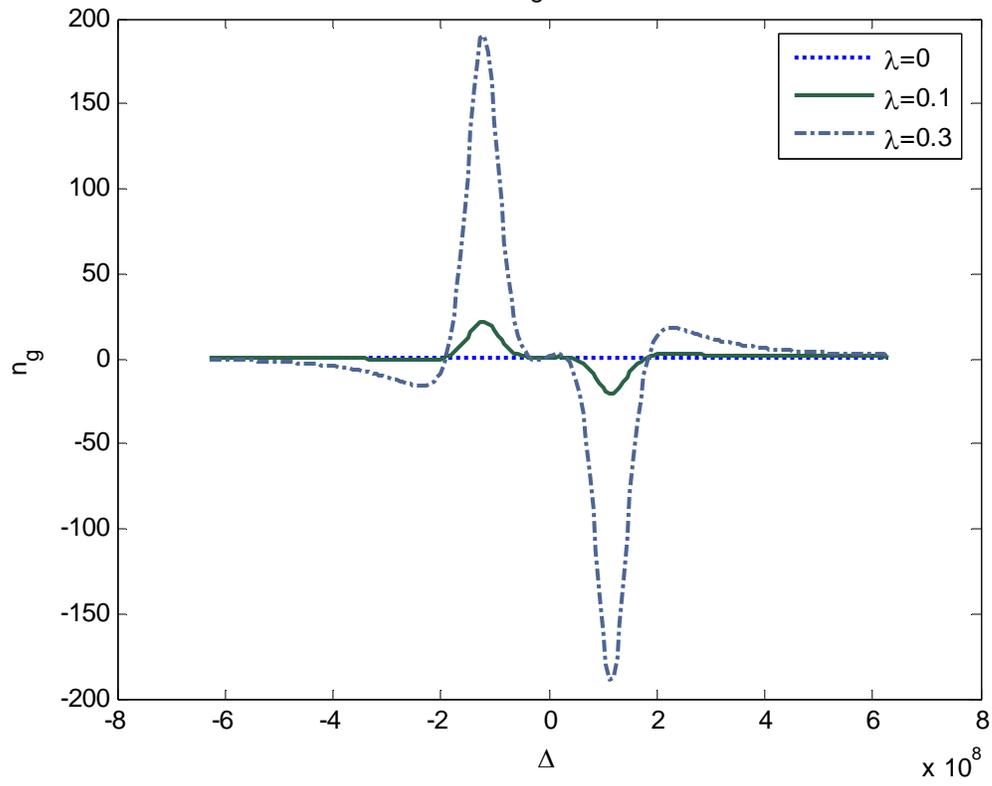

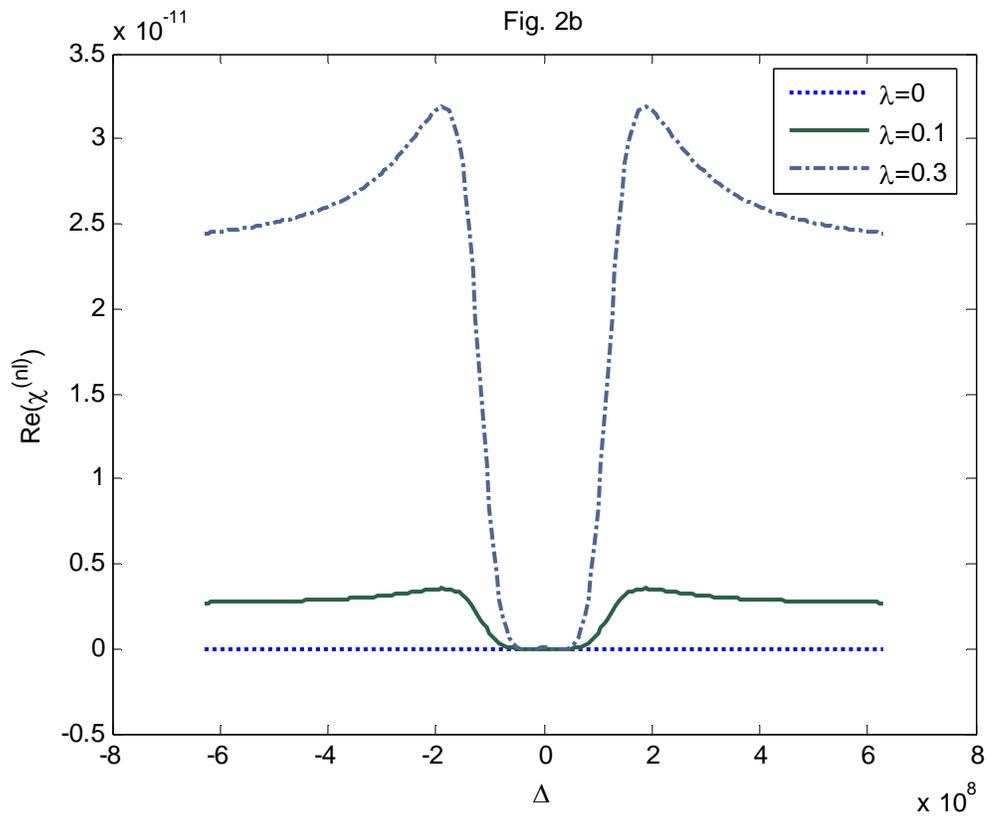

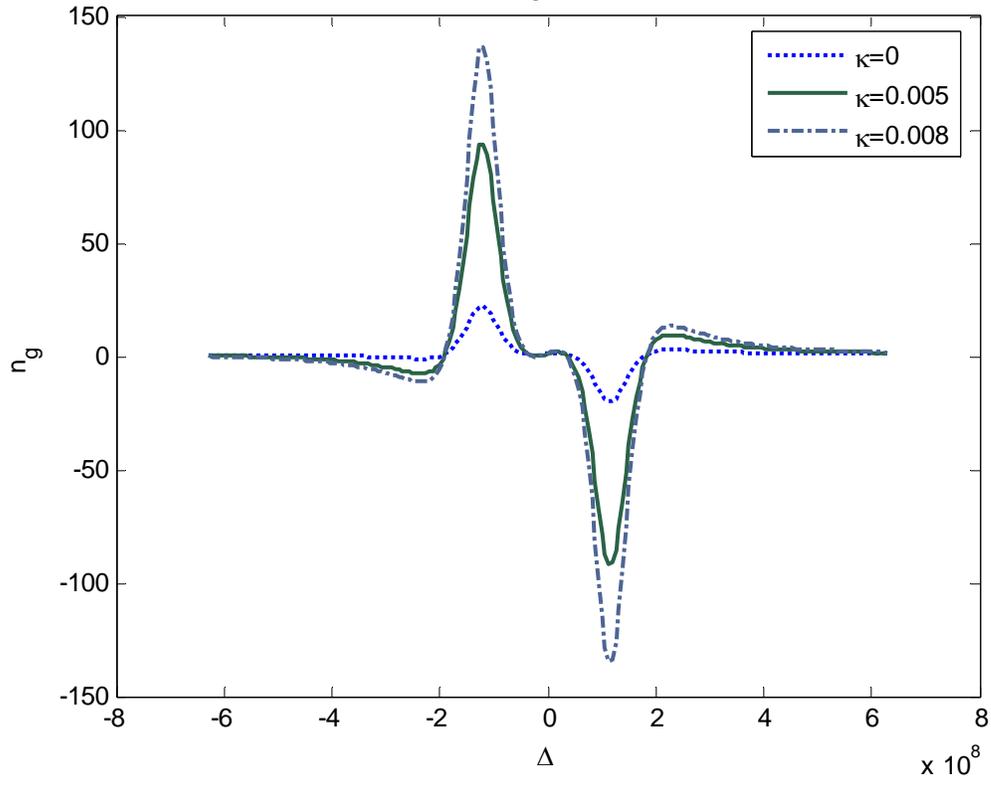

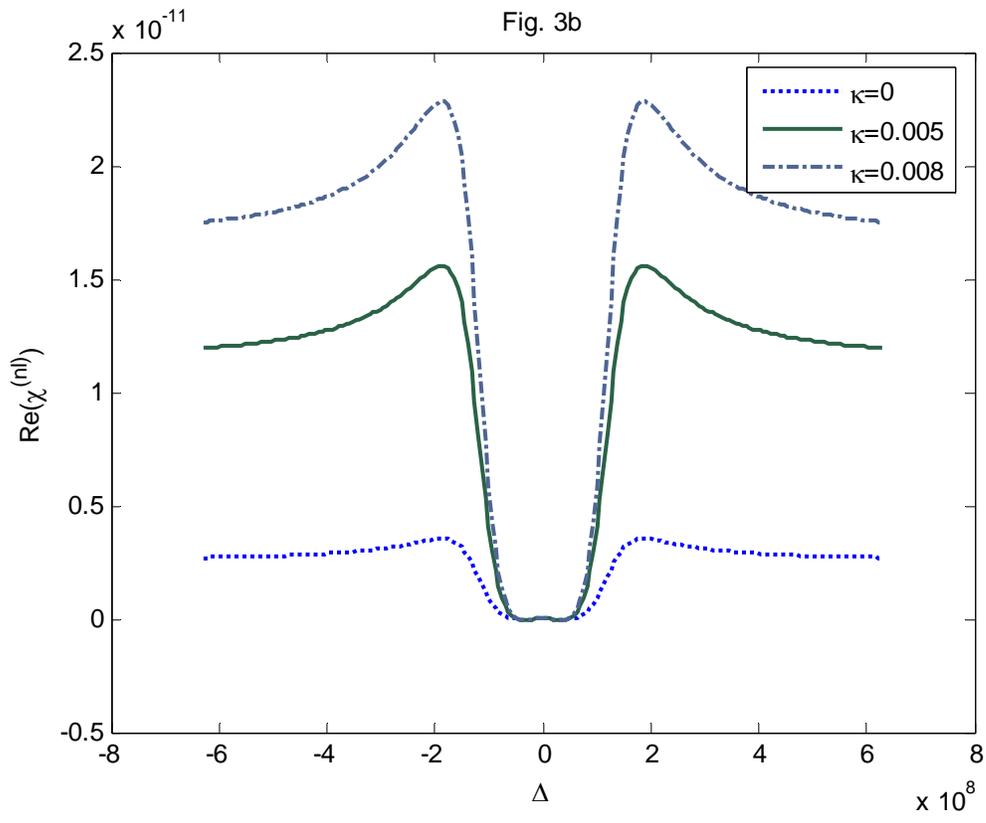

Fig. 3b

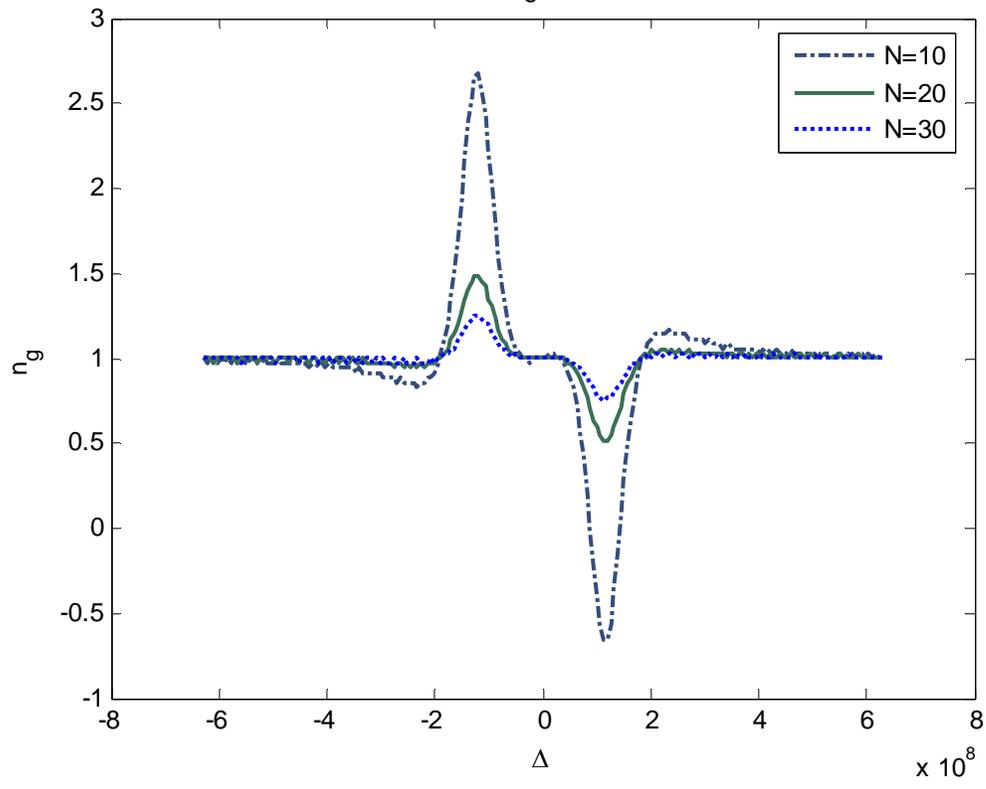

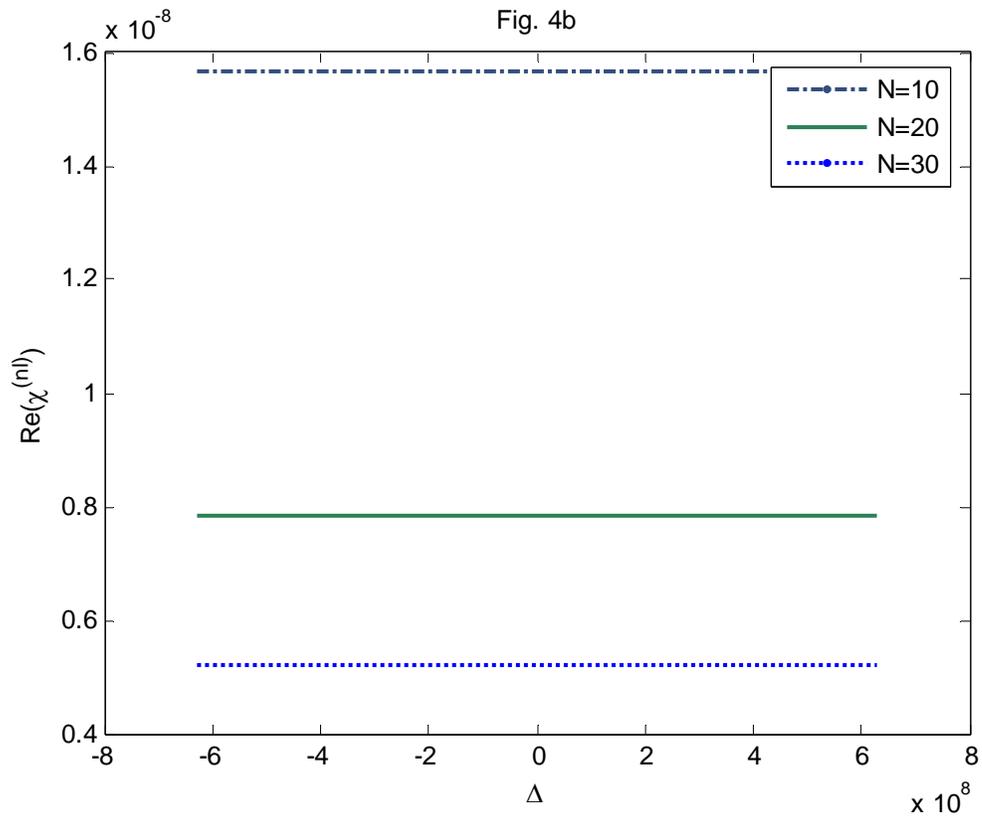

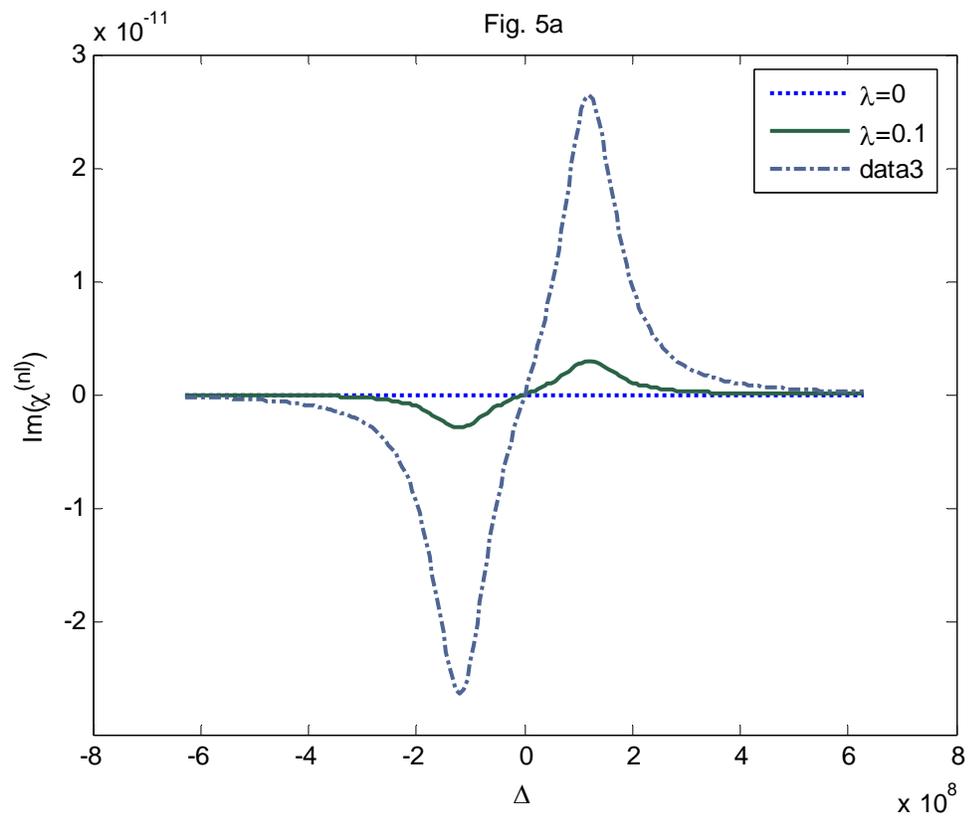

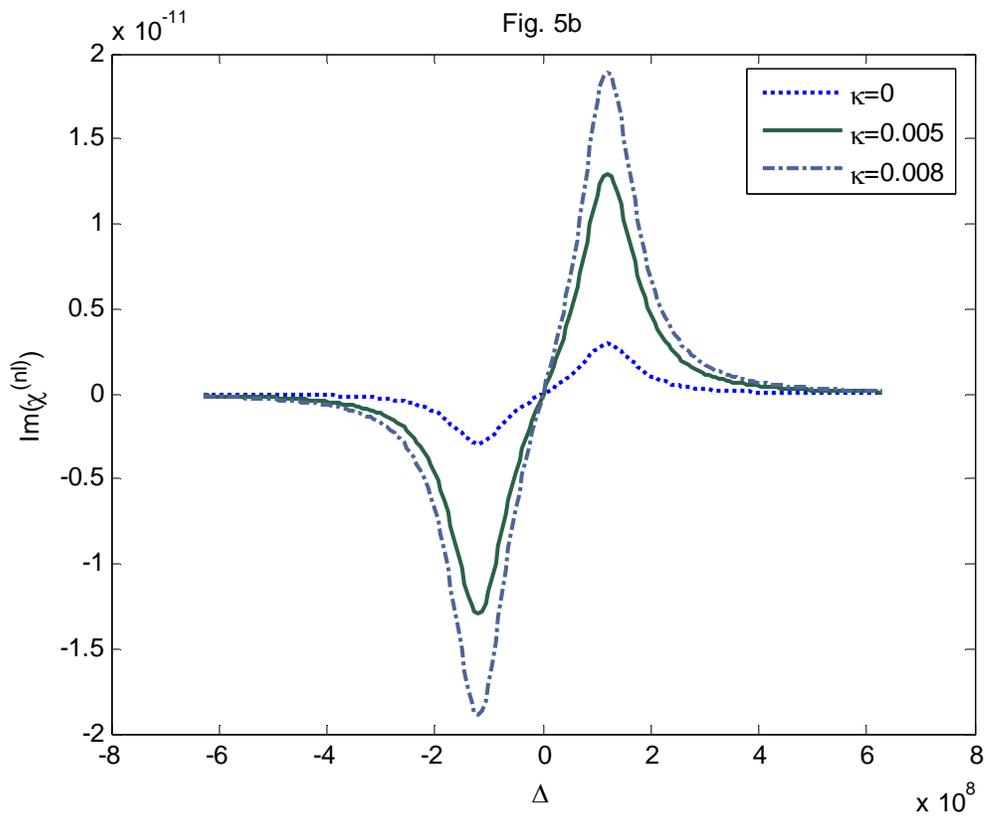

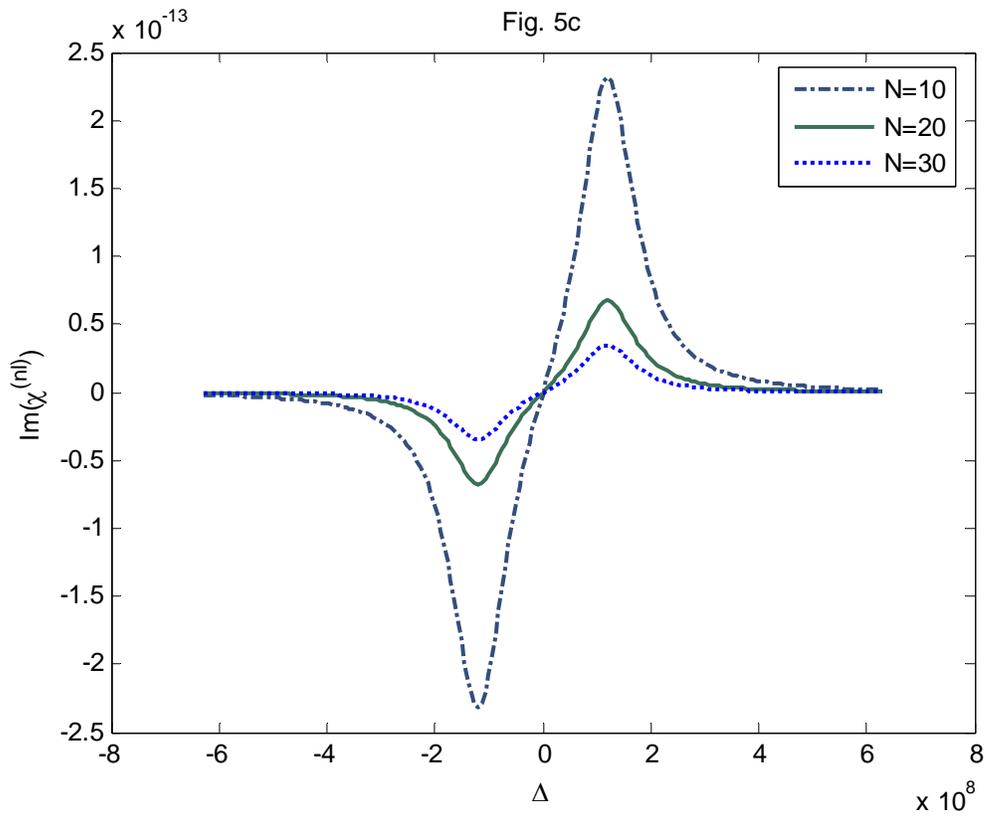

Fig. 5c

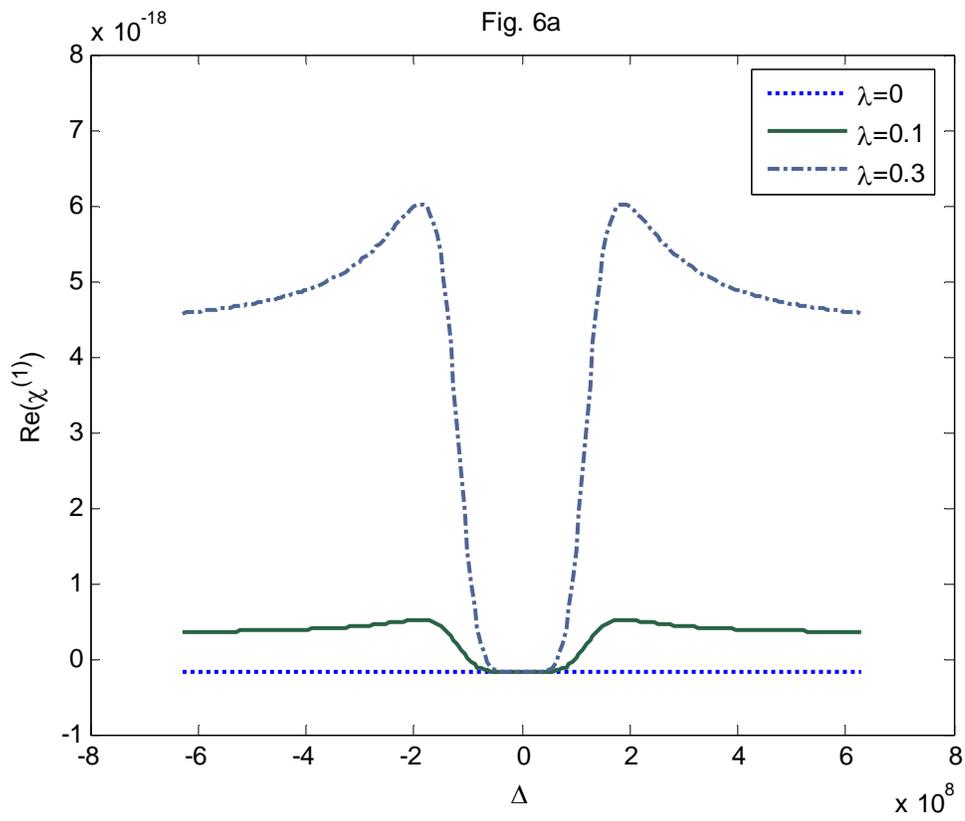

Fig. 6a

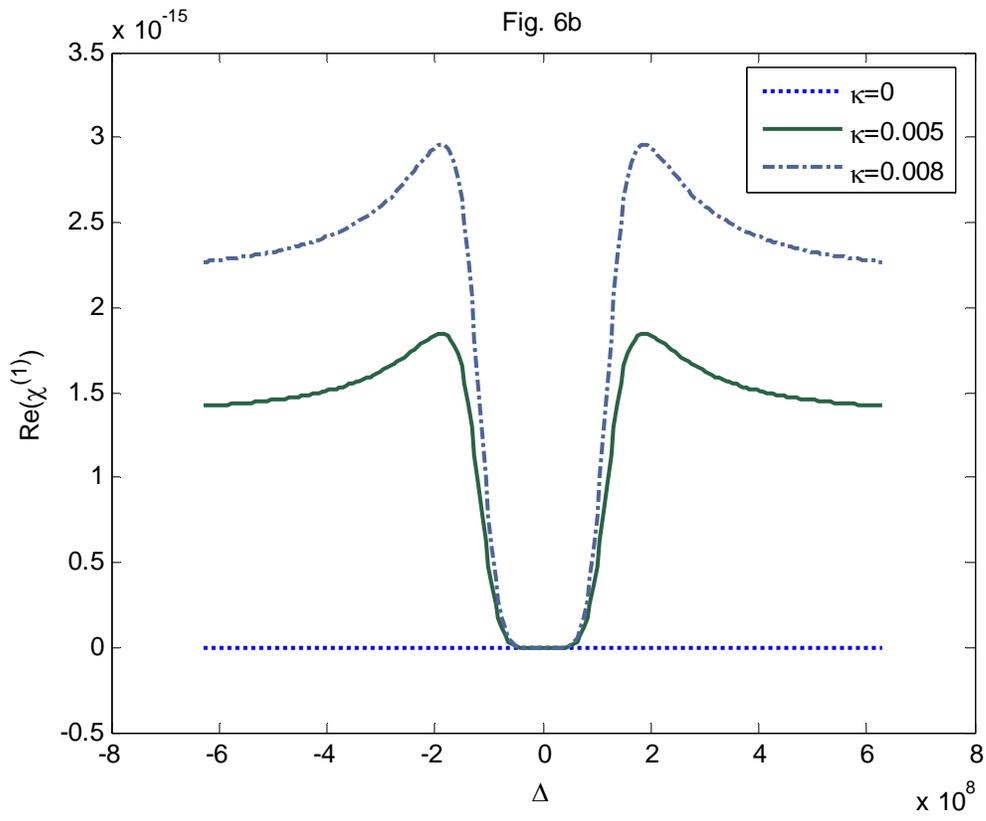

Fig. 6b

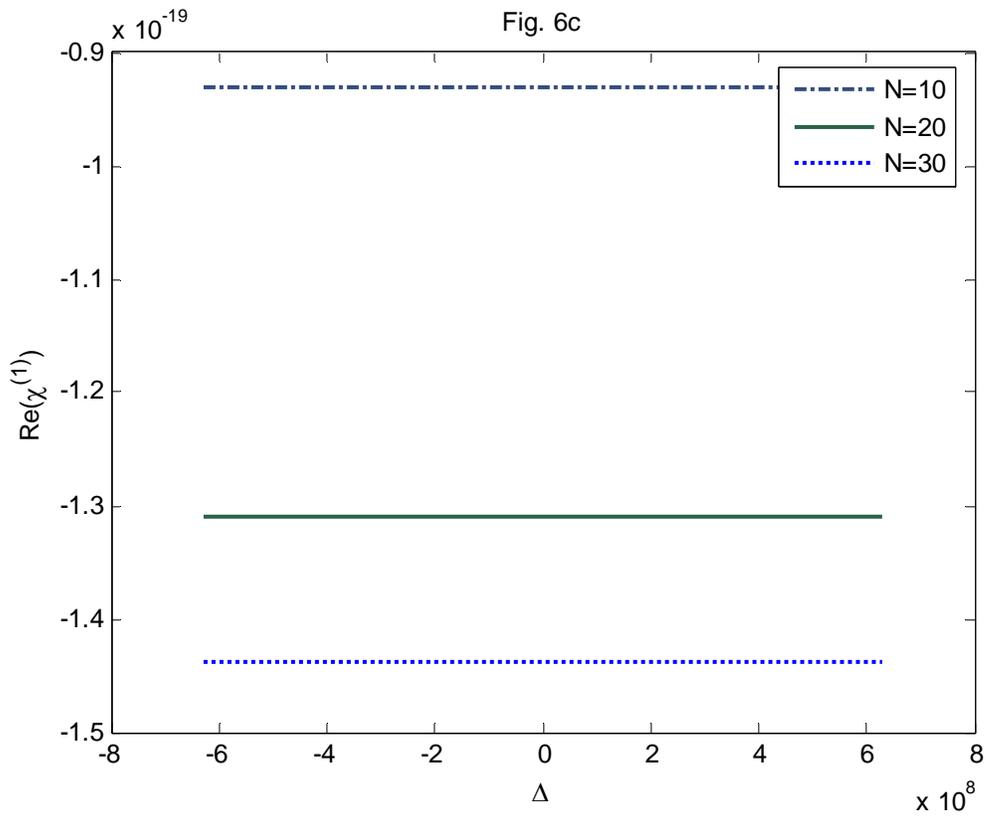

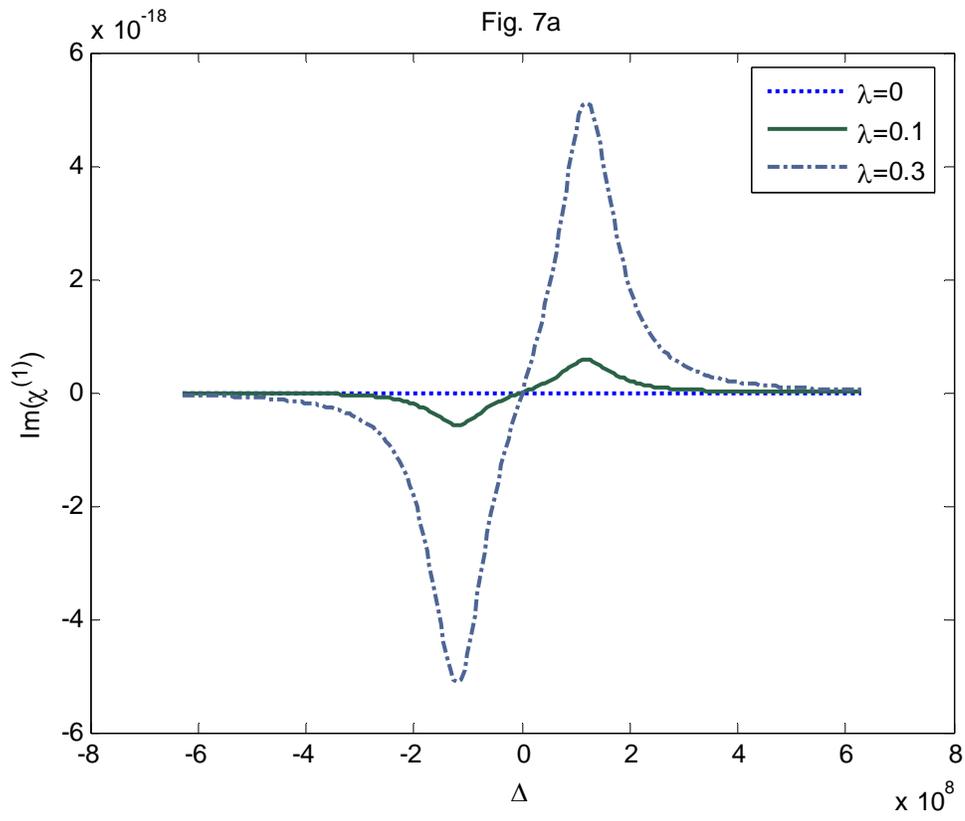

Fig. 7a

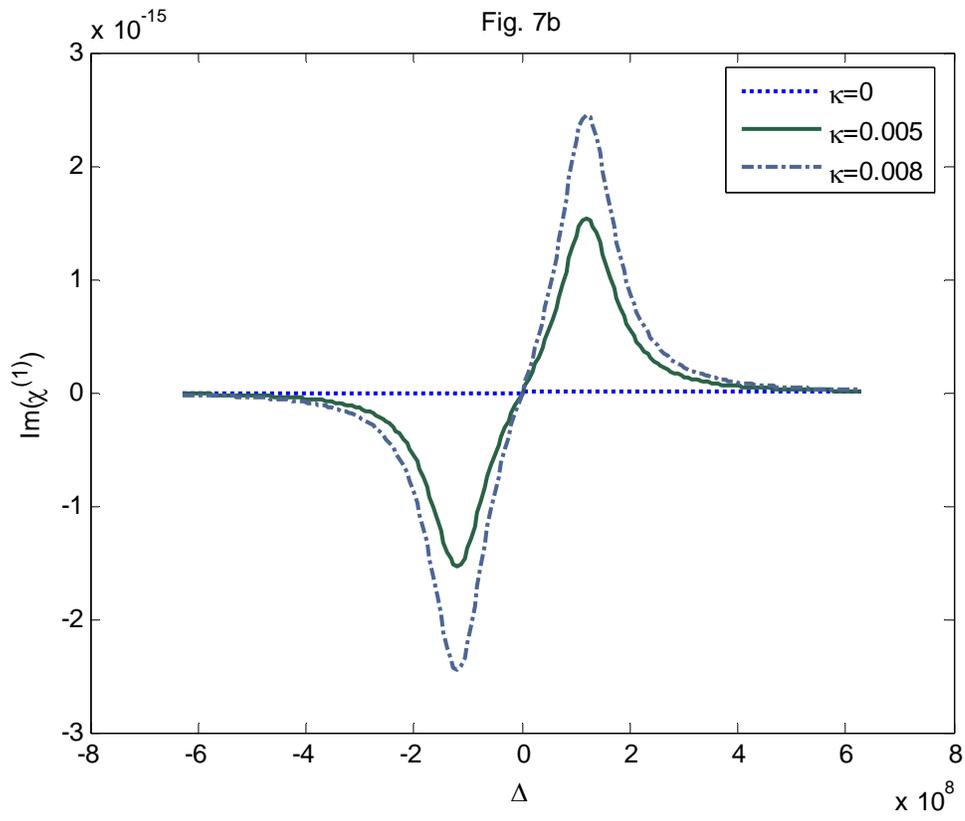

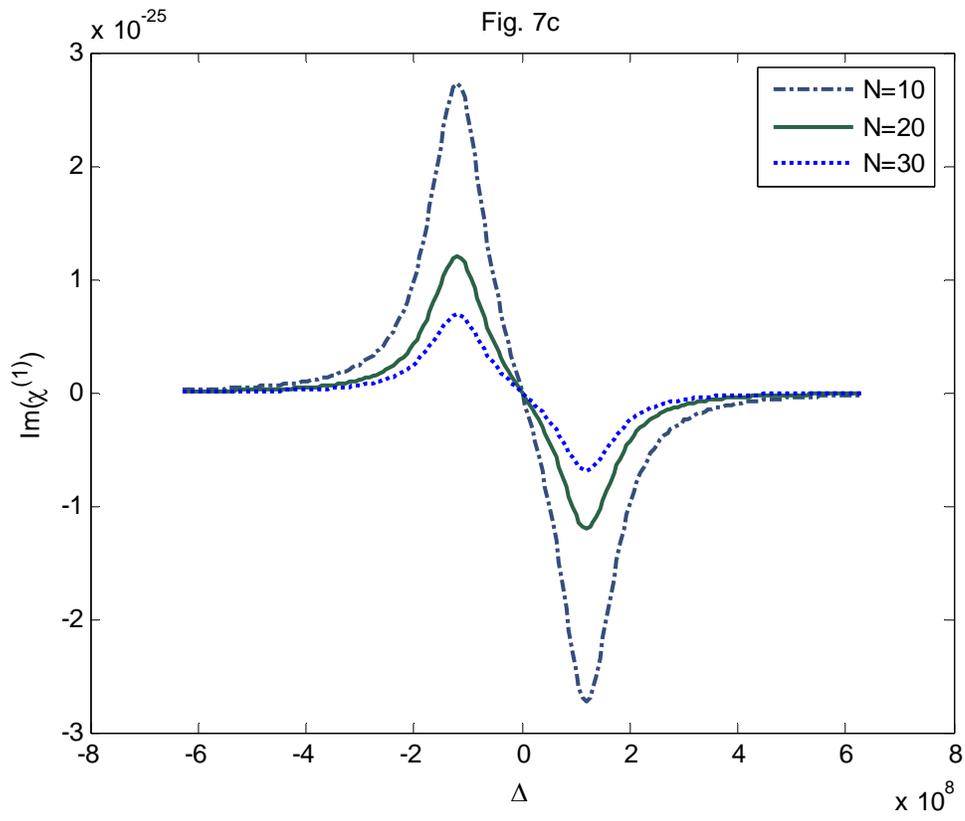

Fig. 7c

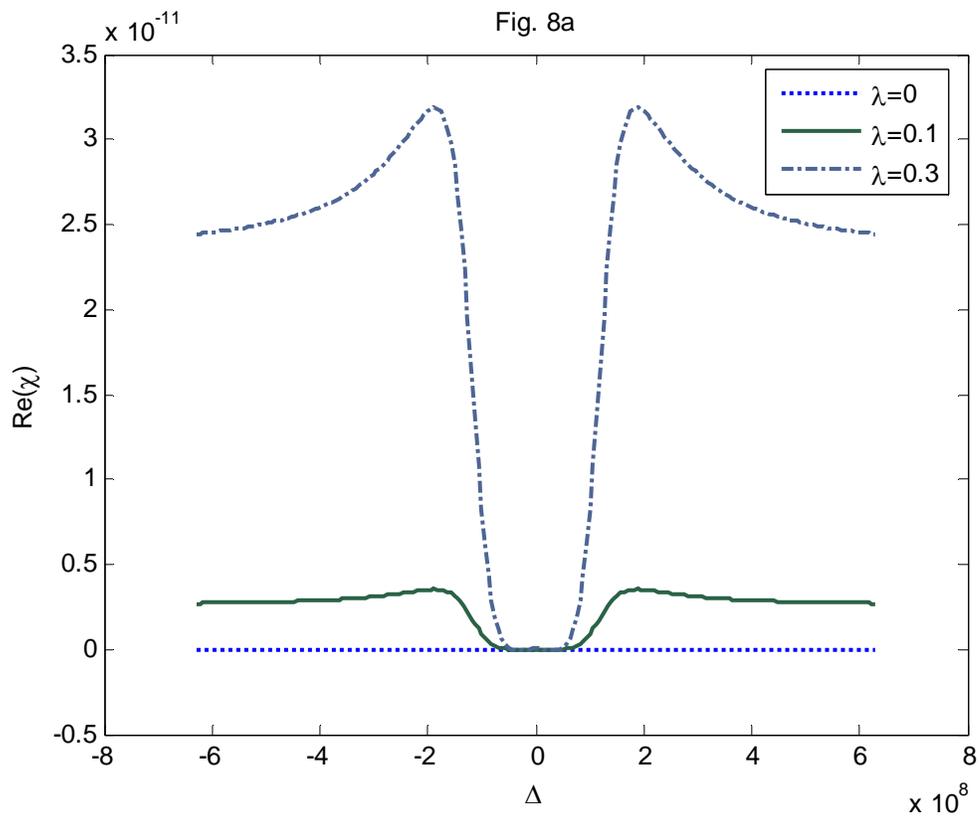

Fig. 8a

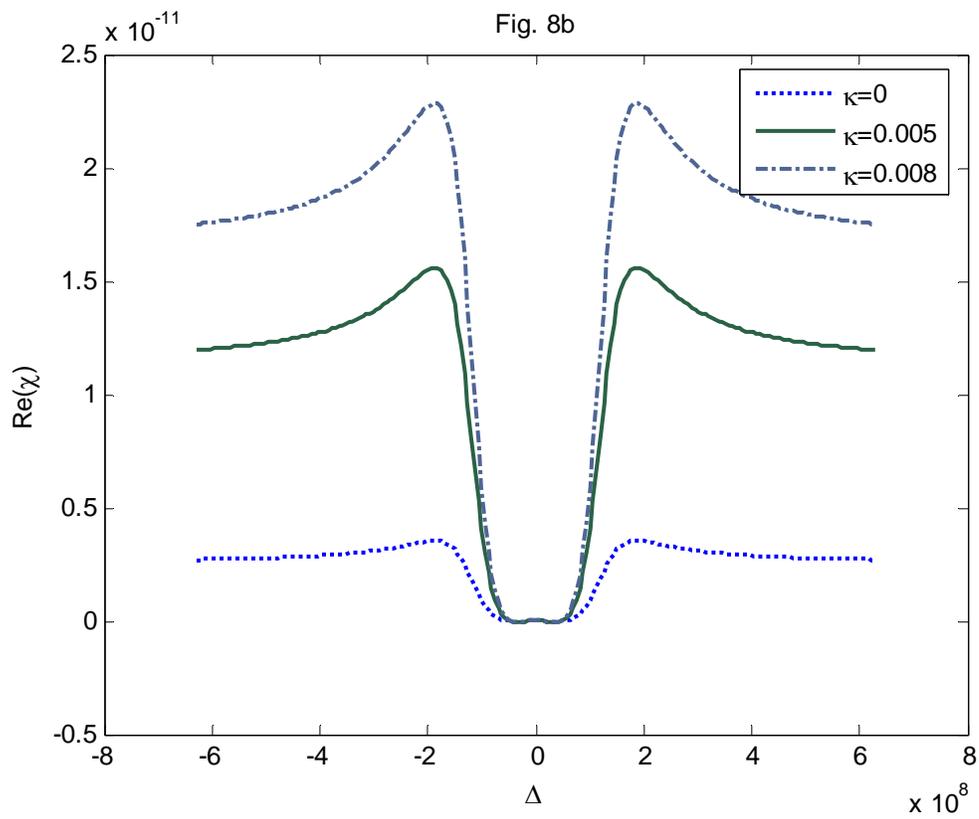

Fig. 8b

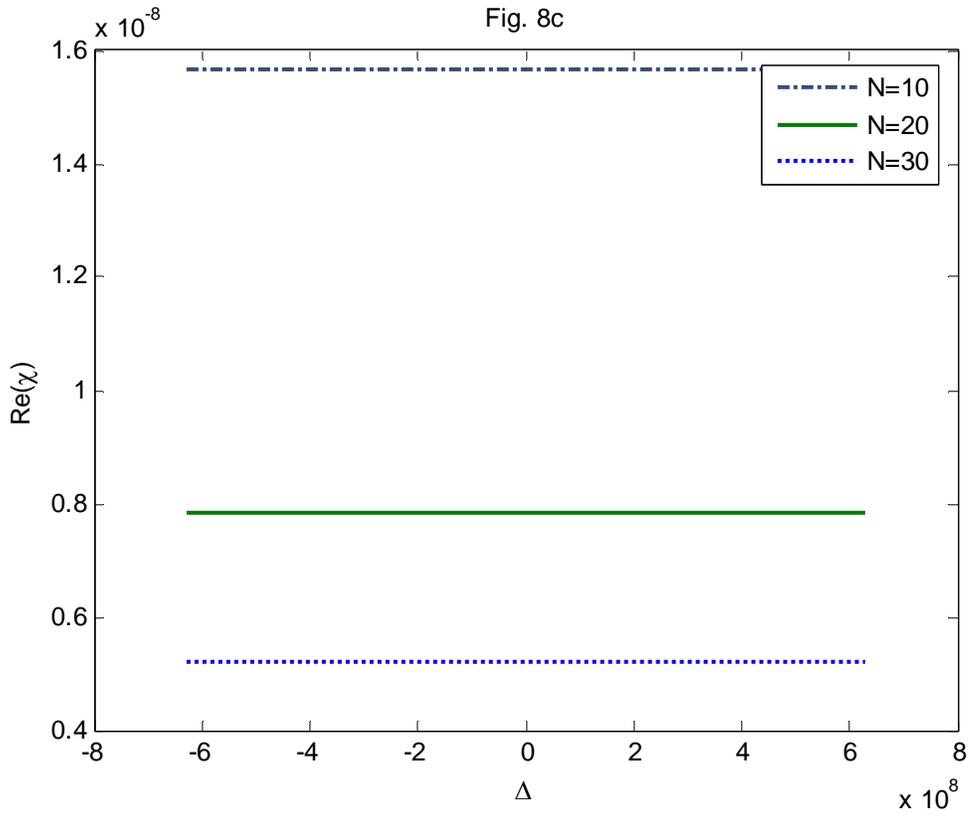

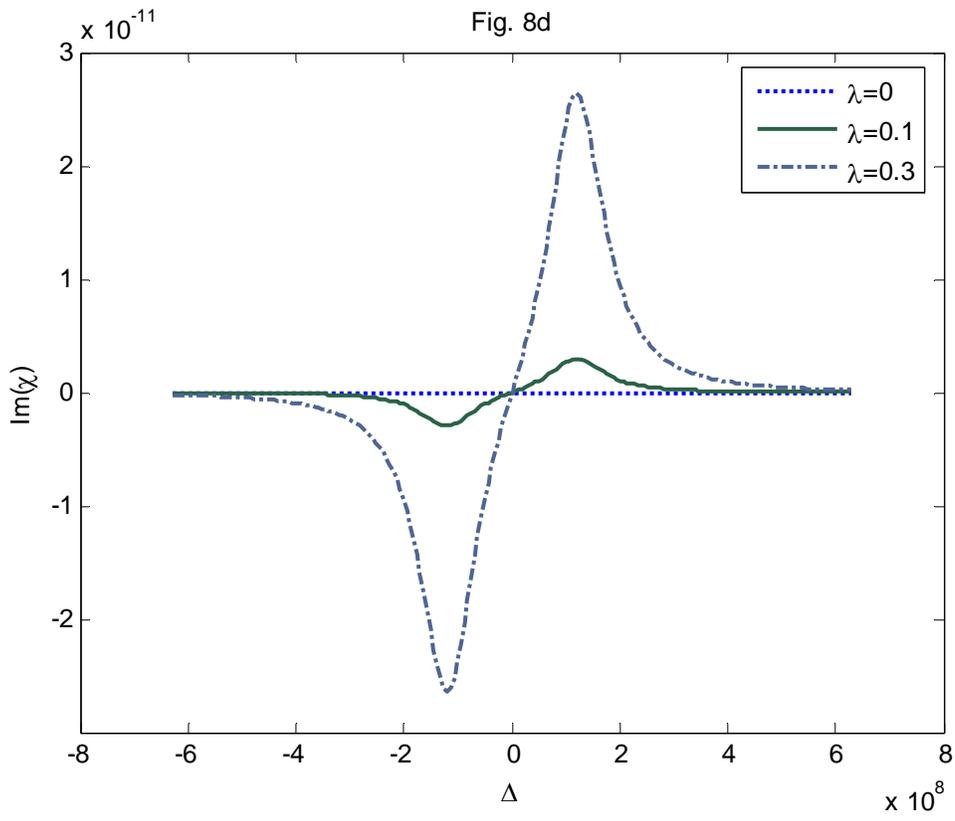

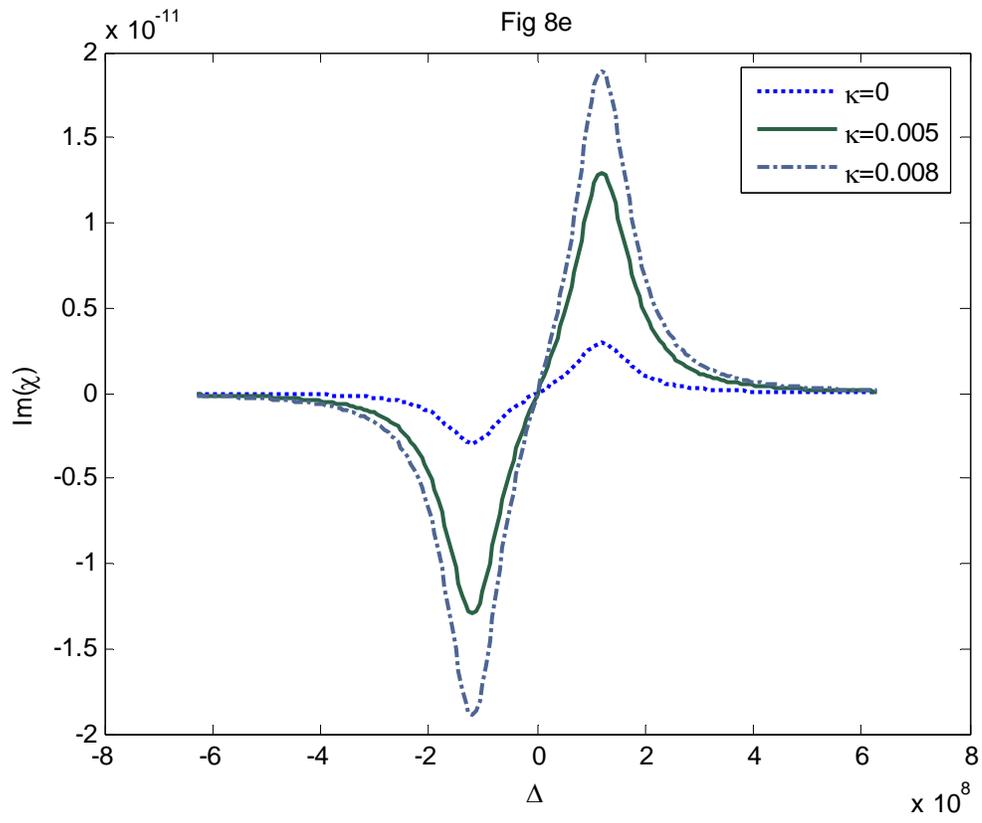

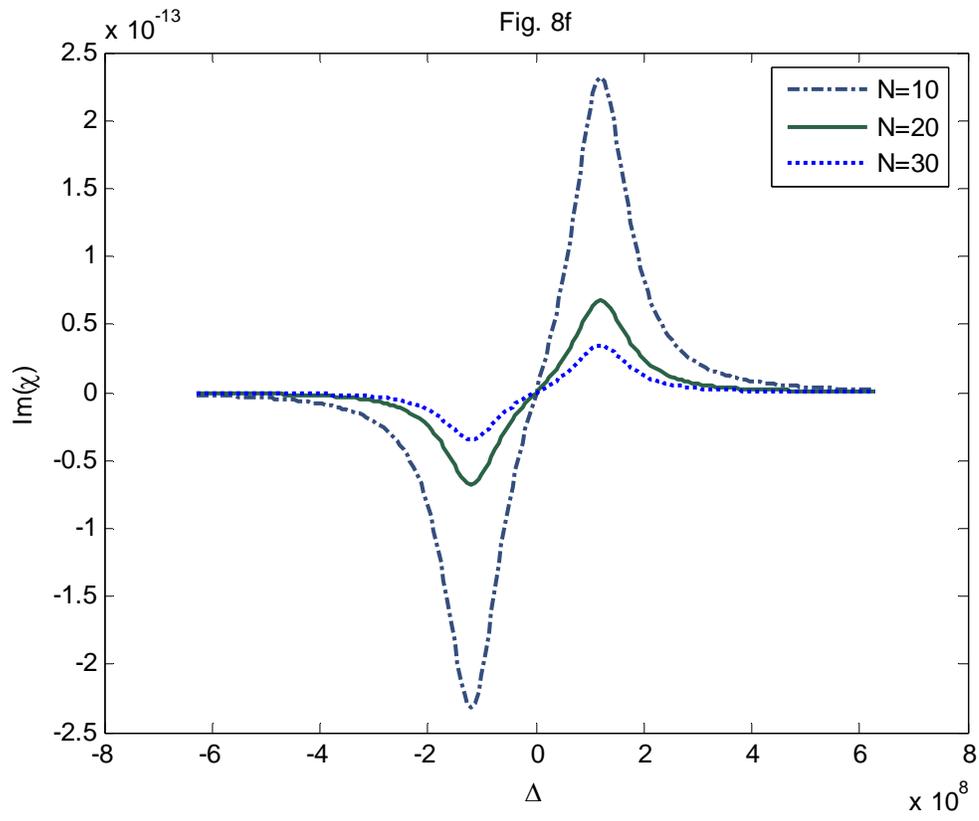

Fig. 8f